\documentclass[a4paper]{article}
\usepackage[]{amsmath,amssymb}
\usepackage{graphics,epsfig}

\begin{document}
\date{}
\title{Entanglement entropy in free quantum field theory}
\author{H. Casini\footnote{e-mail: casini@cab.cnea.gov.ar}\,\, 
and
M. Huerta\footnote{e-mail: marina.huerta@cab.cnea.gov.ar} \\
{\sl Centro At\'omico Bariloche,
8400-S.C. de Bariloche, R\'{\i}o Negro, Argentina}}
\maketitle
\begin{abstract}
In this review we first introduce the general methods to calculate the entanglement entropy for free fields, within the Euclidean and the real time formalisms.  Then we describe the particular examples which have been worked out explicitly in two, three and more dimensions. 
\end{abstract}
\tableofcontents
\newpage
\section{Introduction}

The entanglement entropy associated to a region $V$ of the space in quantum field theory (QFT), is a quantity with manifold interests, ranging from the area of quantum computing and the study of quantum systems in terms of their information content \cite{chuang,upper}, to the physics of black holes \cite{bombelli,cw,srednicki} and the holographic principle \cite{holo}, passing through investigations with interest intrinsic to condensed matter \cite{cm,ami,vidal} and QFT itself \cite{cc,qft,ch1}. 

The problem of an observer who has access only to a subset of the complete set of observables associated to a quantum system gives us the precise scenario to introduce the concept of the entanglement entropy. 
The trace over the degrees of freedom localized on a region which is not accessible to the observer, results in a reduced density matrix. In particular, if the state of the quantum system is the vacuum,  
the local density matrix $\rho_V$ reduced to a region $V$ of the space is
\begin{equation}
\rho_V=\textrm{tr}_{-V}\left|0\right\rangle\left\langle 0\right|\,,
\end{equation}
where the trace is over the degrees of freedom lying outside $V$. Then, the entanglement entropy is the von Neumann entropy associated to $\rho_V$
\begin{equation}
S(V)=-\textrm{tr}(\rho_V\log \rho_V)\,.
\end{equation}

The free quantum field theories play here, as is often the case, the role of the simplest models available to the research. However, even in the free case exact results on the entanglement entropy are scarce, and require  solving difficult questions about the spectrum of certain integral operators, or equivalent questions expressed as functional integrals. Powerful numerical methods on the lattice are known which reduce the calculation of $S(V)$ to a diagonalization of few $\sim m\times m$ matrices, where $m$ is the number of lattice points in $V$. 
This is in contrast to the $\sim q^m\times q^m$ matrix which is at least expected in a generic model with $q$ states per site. This important reduction in computing time allows one to take a close look at the behavior of the entropy function in the free case, at least for low dimension space-times. 
 
The difference between free and interacting QFT models is not evident in terms of the entanglement entropy alone, and has not been elucidated yet in these terms. This makes the study of the free models more interesting because it may reveal features which are common to all QFT. In particular, the ultraviolet behavior is one of these features. In $d$ spatial dimensions we should have for any QFT 
 \begin{equation}
 S(V)=g_{d-1}[\partial V]\,\epsilon^{-(d-1)}+...+ g_1[\partial V]\,\epsilon^{-1} + g_0[\partial V]\,\log (\epsilon)+ S_0(V)\,,   \label{div}
 \end{equation}
where $S_0(V)$ is a finite part, $\epsilon$ is a short distance cutoff, and the $g_i$ are local and extensive  functions on the boundary $\partial V$, which are homogeneous of degree $i$. The leading divergent term coefficient  $g_{d-1}[\partial V]$ is then proportional to the $(d-1)$ power of the size of $V$, and this is usually referred to as the area law for the entanglement entropy. 

This area law is a consequence of the large number of degrees of freedom at high energies present in the QFT which induce entanglement across the boundary $\partial V$. The divergent terms are essentially produced by high energy vacuum fluctuations and thus they are the same for all finite energy density states, such as a thermal state.\footnote{There is a different meaning of area law which is used is the literature of discrete systems which in fact concerns the behavior of the entropy in the large volume limit (rather than a fixed volume in the continuum limit as in (\ref{div})) for specific states such as the ground state, or a state with chemical potential (for a review and a complete list of references see \cite{ecp}). It is known that for fermions in a lattice and finite chemical potential the entanglement entropy grows as the area with a multiplicative logarithmic correction \cite{fermionic1}. However, this does not affect the formula (\ref{div}) since these terms induced by the chemical potential are finite and independent of the cutoff in the continuum limit, that is, they are included in the finite part $S_0$. The same holds for the finite volume increasing terms in a thermal state, which should also satisfy an area law in the sense of (\ref{div}).}

These terms proportional to $g_i$ for $i> 0$ are not physical within QFT since they are not related to quantities well defined in the continuum. On the contrary, the coefficient $g_0$ of the log term is expected to be universal (in this review we use the word universal in the sense of independence of the regularization prescription or of the microscopic model used to obtain the continuum QFT at distances large with respect to the cutoff). There are also many universal terms which are included in $S_0$, but have to be extracted taking into account that the definition of $S_0$ is affected by finite changes in the cutoff. 
 
\begin{figure} [tbp]
\leavevmode
\centering
\epsfxsize=15.8cm
\bigskip
\epsfbox{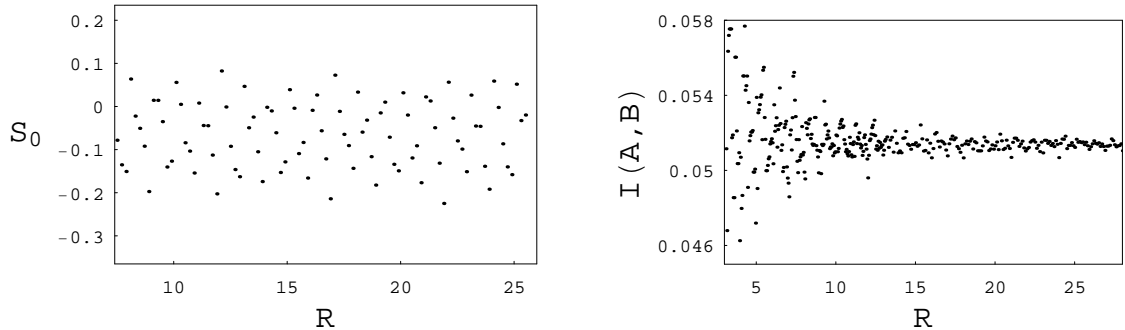}
\caption{The figure on the left shows the term $S_0$ in the entropy of circles in two dimensions as a function of the  radius $R$, with a square lattice regularization. $S_0$ is obtained by fitting the data as $S(R)=c_1 R+ S_0$, and subtracting the term $c_1 R$. It is apparent that $S_0$ does not converge to any definite value. The figure on the right shows the mutual information $I(A,B)$ between two circles $A$ and $B$ of radius $R$ on the lattice, separated by a distance to each other which we have chosen to be also $R$. The calculation shows the convergence of the mutual information in the continuum limit $R\rightarrow\infty$ (with an error of $\sim 2\%$ already for $R\sim 30$). In both figures we have considered a massless scalar field and  $R$ is measured in lattice units. }
\label{figu0}
\end{figure}

Among the universal quantities related to the entropy an important role is played by the mutual information $I(A,B)$ between two non intersecting regions $A$ and $B$ \cite{chuang}, (see figure \ref{figu0})
\begin{equation}
I(A,B)=S(A)+S(B)-S(A\cup B)\,.
\label{mutual}
\end{equation}
The universal character is due to the cancellation of the boundary terms. 
Another useful dimensionless universal quantity defined in two dimensions and for an interval of size $L$ is given by  
\begin{equation}
c(L)=L \frac{dS(L)}{dL}
\,. \label{centralcharge}
\end{equation}
The entropic $c$-function $c(L)$ is always positive and decreasing, and plays the role of the Zamolodchikov's $c$-function \cite{cteor} in the entanglement entropy $c$-theorem \cite{ch1}. It contains all the universal information present on the entropy for an interval, since $S(L)$ follows from $c(L)$ by integration, except for an arbitrary constant. Accordingly, in this review we will present the results for the entanglement entropy for single intervals in two dimensions in terms of $c(L)$.

Still related to the local density matrix, there is a family of functions called the alpha or Renyi entropies,  
\begin{equation}
S_n(V)=\frac{1}{1-n}\log \textrm{tr}\rho^n_V\,,\label{esealfa}
\end{equation}
which share some of the entanglement entropy properties.
These are often easier to compute, and one has
\begin{equation}
\lim_{n\rightarrow 1} S_n(V) =S(V)\,.\label{alpha}
\end{equation}
For an interval in two dimensions a quantity analogous to the entropic c-function (\ref{centralcharge}) can be defined as
\begin{equation}
c_n(L)=L \frac{dS_n(L)}{dL}\,, \hspace{1cm} \lim_{n\rightarrow 1} c_n(L)=c(L)\,.\label{charge1}
\end{equation}

This review is focused on vacuum entanglement entropy in relativistic free QFT and its associated universal quantities.  In what follows we describe two general methods to calculate the geometric entropy for free fields. We then give examples where these methods are applied in two, three and more dimensions for scalar and Dirac fields. Free gauge fields have not been sufficiently studied in the literature yet. They have an additional technical problem since the divergent terms in (\ref{div}) are non gauge invariant. 

Most of the material is already present in the literature, but we have also included some new results, and reformulated earlier work.  
There are several topics related to entanglement for free fields discussed in the literature but not belonging to our main subject which will not be covered here. These topics include entanglement entropy in presence of  temperature, chemical potential, or excited states \cite{cc,fermionic1,kor,otrosfermionic}, manifolds with boundary \cite{frolovangulo,dc,sh,met}, different (Newton-Wigner) localization schemes \cite{nw}, entanglement in curved space-time \cite{dos,ii}, and in non-Lorentz covariant QFT (quantum Lifshitz fixed points) \cite{frad}. There is also a large body of work done in the context of lattice models, which extends out of the scope of this review. There are several excellent review papers in this area \cite{ami,ecp,pesc}.   

\section{The entanglement entropy in free QFT}
\subsection{Euclidean time method}
\begin{figure} [tbp]
\centering
\leavevmode
\epsfysize=4cm
\bigskip
\epsfbox{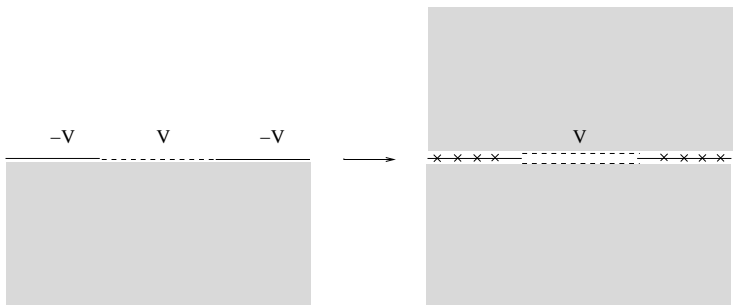}
\caption{The path integral on the lower half Euclidean space gives the vacuum wave functional. The reduced density matrix $\rho_V$ is obtained by gluing two copies of this half space along the set $-V$ complementary to $V$.}
\label{figu1}
\end{figure}

In the Euclidean approach one starts from the representation of the vacuum state in terms of a path integral.  For definiteness consider a scalar field $\hat{\phi}(\vec{x},t)$, and take the basis formed by eigenvectors of this field operator at time $t=0$ as $\hat{\phi} (\vec{x},0) \left|  \alpha\right>= \alpha(\vec{x})\left|  \alpha\right>$, where $\alpha$ is any real function on the space. The vacuum wave functional writes \cite{book}  
\begin{equation}
\Phi(\alpha)=\left<0|\alpha\right>=N^{-1/2}\int_{\phi(\vec{x},-\infty)= 0}^{\phi(\vec{x},0)=\alpha(\vec{x})} D\phi\,\, e^{-S_E(\phi)}\,. \label{qs}
\end{equation} 
In order to select the vacuum state, the functional integral is over the lower half space and with Euclidean time. $S_E(\phi)$ is the Euclidean action and $N^{-1/2}$ is a normalization factor. The vacuum density matrix in this basis is
$\rho(\alpha,\alpha^\prime)=\left<\alpha | 0\right> \left< 0| \alpha^\prime\right>=\Phi(\alpha)^*\Phi(\alpha^\prime)$. In order to trace over degrees of freedom in $-V$, the set complementary to $V$, one considers functions $\alpha=\beta \oplus \alpha_{V}$, $\alpha^\prime=\beta\oplus \alpha^\prime_{V}$, which coincide (are equal to $\beta$) on $-V$, and sum over all possible functions $\beta$. Using the representation (\ref{qs}) this construction of the reduced density matrix amounts to take two copies of the half space, glue them on $-V$ (see figure \ref{figu1}), and take the functional integral in this space \cite{ci},
\begin{equation}
\rho_V(\alpha_{V},\alpha^\prime_{V})=\int D\beta\,\, \Phi(\beta \oplus \alpha_{V})^* \Phi(\beta\oplus \alpha^\prime_{V})= N^{-1}\int_{\phi(\vec{x},0^-)=\alpha^\prime_V(\vec{x}),\, x\in V }^{\phi(\vec{x},0^+)=\alpha_V(\vec{x}),\,x\in V} D\phi\,\, e^{-S_E(\phi)}\,. \label{opre}
\end{equation}
The arguments of the density matrix are the boundary conditions of the path integral on both sides of the cut on $V$. 

For fermions a similar construction holds, but there is an important difference. Since the fields at equal time anticommute, the functional integral and the boundary conditions are in terms of  Grassmann variables. In order to represent the trace in the functional integral we have to sum over the antidiagonal elements \cite{klein}, that is  
\begin{equation}
\rho_V(\alpha_{V},\alpha^\prime_{V})=\int D\beta\,\, \Phi(-\beta \oplus \alpha_{V})^* \Phi(\beta\oplus \alpha^\prime_{V})\,,
\end{equation}
where now all variables involved are Grassmann valued. This extra minus sign is well known in the calculation of the thermal partition function for fermions, which requires antiperiodic boundary conditions.  Here, this implies that when we glue the two copies of the half space along $-V$ we have to take boundary conditions related by a minus sign. This cut at $-V$ can be eliminated by changing integration variables $\psi\rightarrow -\psi$ in the upper half plane. The net result is that there is an additional minus sign in the boundary condition for the density matrix in the upper side of the cut over $V$ in figure \ref{figu1},  
\begin{equation}
\rho_V(\alpha_{V},\alpha^\prime_{V})=N^{-1}\int_{\psi(\vec{x},0^-)= \alpha^\prime_V(\vec{x}),\, x\in V }^{\psi(\vec{x},0^+)=-\alpha_V(\vec{x}),\,x\in V} D\psi\,D\bar{\psi}\, e^{-S_E(\psi,\bar{\psi})}\,. 
\end{equation}

The representation of the traces $\textrm{tr}\rho_V^n$ involved in (\ref{esealfa}) by a functional integral is then realized by the replication method which consists on taking $n$ copies of the Euclidean plane cut along $V$, and sewing together the upper side of the cut in the $k^{th}$ copy with
the lower one of the $(k+1)^{th}$ copy, for $ k=1,...,n$ \cite{cw,cc}. Here the copy $n+1$ coincides with the first one (see figure \ref{figu2}). The resulting space is a $n$-sheeted $d+1$ dimensional Euclidean space with conical singularities of angle $2\pi n$ located at the boundary $\partial V$ of the set $V$. For fermions, the final trace (after the power $\rho^n$ has been taken) also involves a sign change in the boundary conditions. We can change variables in order to have all gluing with no additional signs except for the last one, where the field on the upper side of the $n^{th}$ cut is equated to $(-1)^{n+1}$ times the field on the lower side of the first cut \cite{fermion}. Here, one minus sign counts for each copy of the density matrix plus one extra minus sign for the trace\footnote{For the special case of the Rindler space, corresponding to $V$ half a spatial hyperplane, the density matrix is proportional to $e^{-2\pi K}$, with $K$ the boost operator which keeps the Rindler wedge fixed \cite{boostwedge}. In the Euclidean version, the boost operator corresponds to a rotation operator and  $\rho^n=e^{-2\pi n K}$ to a rotation operator of angle $2 \pi n$. When translated into the path integral formalism this provides a different look to the origin of the signs on the fermion boundary conditions in this case \cite{Kabat}.}. 

Finally we have  
\begin{eqnarray}
\textrm{tr}\rho_V^{n}&=& \frac{Z(n)}{Z(1)^{n}}\,,\\
S_n(V)&=& \frac{\log Z(n)-n \log Z(1)}{1-n}\label{cheto}\,,
\label{dd}
\end{eqnarray}
where $Z(n)$ is the functional integral on the $n$-sheeted manifold, and we have used the normalization factor $N=Z(1)$ in order to have $\textrm{tr} \rho_V=1$. 
Eq. (\ref{cheto}) gives a representation of the Renyi entropies for integer $n$. The entanglement entropy follows by analytic continuation of $S_n$ down to $n=1$ (\ref{alpha}).
\subsubsection{Diagonalization in replica space}
In general, calculating the integrals $Z(n)$ explicitly is a very difficult problem since we have to deal with a non trivial manifold resulting from the replication method.
Fortunately, in the case of free fields, a simplification  follows by mapping the $n$-sheeted problem to an equivalent one in which one deals with $n$ decoupled and multivalued free fields  \cite{fermion}. For that, one introduces a vector field $\vec{\Phi}$ living on a single $d+1$ dimensional space, whose components are the values of the fields in the different copies, 
\begin{equation}
\vec{\Phi}=\left(\begin{array}{c}
\phi _{1}(x)\\
\vdots \\  
\phi _{n}(x) \end{array}
\right) \,,
\end{equation}
where $\phi _{l}(x)$ is the field on the $l^{\textrm{th}}$ copy. 
Note that in this way the space is simply connected but the singularities at the boundaries of $V$ are still there since the vector $\vec{\Phi}$ is not singled valued.  In fact, crossing $V$ from above or from below, the field gets multiplied by a matrix $T$ or $T^{-1}$ respectively, where
\begin{equation}
\begin{array}{c}
T=\left(
\begin{array}{lllll}
0 & 1 &  &  &  \\
& 0 & 1 &  &  \\
&  & . & . &  \\
&  &  & 0 & 1 \\
(\pm1)^{n+1} &  &  &  & 0
\end{array}
\right)
\end{array}\,. 
\end{equation}
The upper sign in this equation corresponds to the bosonic case and the lower one to fermions. 

This matrix has eigenvalues $e^{i\frac{k}{n}2\pi }$, with
$k=0\,,...,\,(n-1)$ (the $n^{\textrm{th}}$ roots of $1$) in the scalar case, and $e^{i\frac{k}{n}2\pi }$ with $k=-(n-1)/2\,, -(n-1)/2+1,...,\,(n-1)/2$ (the $n^{\textrm{th}}$ roots of $(-1)^{n+1}$), in the fermionic one. 
Then, changing basis by a unitary transformation in the replica space,
we can diagonalize $T$, and the problem is reduced to $n$ decoupled
fields $\tilde{\phi}_{k}$ living on a single $d+1$ dimensional space. At this point is essential the free character of the action, and we also need to deal with complex fields. This last requirement is not a limitation since in order to compute the entanglement entropy we can double the number of real fields and then divide the final result by two. The fields which diagonalize $T$ are multivalued and defined on the Euclidean $d+1$ dimensional space with boundary conditions imposed on the $d$ dimensional set $V$ given by
\begin{equation}
\tilde{\phi}_k(\vec{x},0^+)=e^{i\frac{2 \pi k}{n}}\tilde{\phi}_k(\vec{x},0^-)\,\,\,\,\,\,\,\,\,\,,\,\, \, \vec{x}\in V\,.
\label{bc}
\end{equation}
\begin{figure} [tbp]
\centering
\leavevmode
\epsfxsize=10cm
\bigskip
\epsfbox{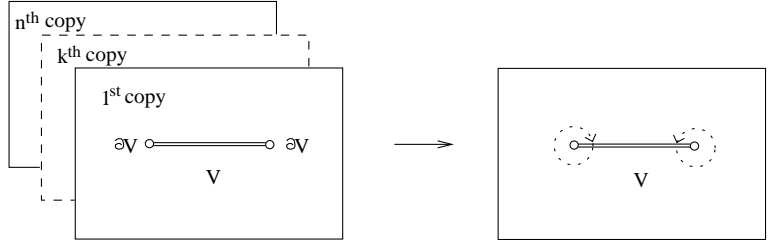}
\caption{tr$\rho^n_V$ is given by the path integral on a $n$-sheeted space formed by sewing the replicated Euclidean spaces with a cut along $V$.  This is equivalent for free fields to $n$ decoupled multivalued fields living in a single space. These fields get multiplies by particular phase factors when crossing the cut.}
\label{figu2}
\end{figure}
Here $\tilde{\phi}_k(\vec{x},0^\pm)$ are the limits of the field as the variable approaches $V$ from each of its two opposite sides in $d+1$ dimensions.
 
In this formulation we have for scalars 
\begin{equation}
S_n(V)=\frac{1}{1-n}\sum_{k=0}^{n-1}\textrm{log}Z[e^{i 2 \pi k/n}]\,,
\label{r1}
\end{equation}
and for fermions 
\begin{equation}
S_n(V)=\frac{1}{1-n}\sum_{k=-(n-1)/2}^{(n-1)/2}\log Z[e^{i 2 \pi k/n}]\,,
\label{r2}
\end{equation}
where $Z[e^{i 2 \pi a}]$ is the partition function corresponding to a field which acquires a phase $e^{i2\pi a}$ when the variable crosses $V$ (figure \ref{figu2}), divided by $Z(1)$. 

To further specify these partition functions one has to take into account that the fields must have a specific asymptotic behavior as the variable approaches the singularity in order to have finite action. This requirement will be different according the action is quadratic (scalars) or linear (fermions) in derivatives. For smooth $V$ boundaries we have 
\begin{eqnarray}
\phi(x) &\sim& r^\gamma\,,\hspace{1.5cm} \gamma > 0 \hspace{1.05cm}\textrm{(scalars)}\,,\label{chia}\\
\psi(x) &\sim& r^\gamma\,, \hspace{1.5cm}  \gamma > -1/2  \hspace{.4cm}\textrm{(fermions)} \,,
\label{choa}\end{eqnarray}
in the limit of short $r$, where $r$ is the distance between $x$ and $\partial V$. 

Note that (\ref{r1}) and (\ref{r2}) give the same formula for the limit $n\rightarrow \infty$,
\begin{equation}
S_\infty(V)=-\frac{1}{2 \pi} \int d\theta\,\, \log Z[e^{i \theta}]\,.
\end{equation}

\subsubsection{Analytic continuation}
The evaluation of the entropy is still limited by the difficulty in doing the analytic continuation of  $S_n$ for non-integer $n$, and then the limit (\ref{alpha}). The analytic continuation of a function defined on the integers is not unique unless some further information is provided. One such requirement would be the condition $S_n<c\, e^{\pi |n| }$ for Re$(n)>1/2$, which is the hypothesis of the Carlson's theorem \cite{ccd,carl}. This holds for finite dimensional density matrices, but for universal terms in $S_n$ in a continuum theory no general result is known. For a discussion around this point see \cite{ccd,d1}. 

The analytic continuation for the free fields which satisfies Carlson's criterion can be obtained  in a natural way writing the sums in $S_n$ as a contour integrals \cite{chana}. We will check in section 2.3 that the result coincides with the entanglement entropy obtained by the real time approach. 

\begin{figure} [tbp]
\centering
\leavevmode
\epsfysize=4.3cm
\bigskip
\epsfbox{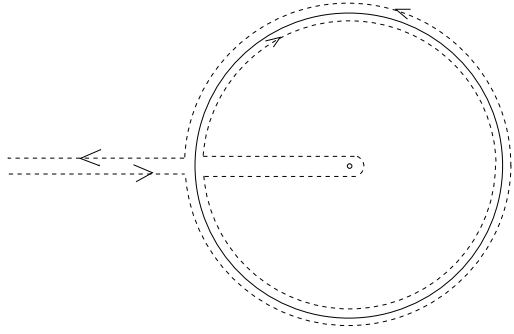}
\caption{The integration path used in eqs. (\ref{chuta}) and (\ref{chuta2}) to define the analytic continuation of $\textrm{tr}\rho^n$ for non-integer $n$.}
\label{figu22}
\end{figure}

We treat first the scalar case. Eq. (\ref{r1}) can be written 
\begin{equation}
S_n=\frac{1}{2\pi i (1-n)}\sum_{k=0}^{n-1}\oint du \, \frac{\textrm{log}Z[u]}{u-e^{i 2 \pi k/n}}=\frac{1}{2\pi i (1-n)}\oint du \,\frac{n\, u^{n-1}}{u^n-1}\,\textrm{log}Z[u]\,,\label{chuta}
\end{equation}
where the contour of integration encircles the roots of the unit. The function $\textrm{log}Z[u]$ is defined as the unique analytic continuation of the function $\log Z[e^{i 2\pi a}]$ on the unit circle.
 Since  we need to avoid the negative real axis for the formula (\ref{chuta}) to be analytic for non integer $n$, we choose the integration contour as in figure \ref{figu22}. Thus, we are assuming $\textrm{log}Z[u]$ is analytic on the negative real axes and $\left|\textrm{log} Z[u]\right|/\left|u\right|^x\rightarrow 0$ as $u\rightarrow-\infty$ for some $x<1$ (we will show this holds for the universal terms in $\log Z[u]$ in section 2.3). In the limit $n\rightarrow 1$ we have
\begin{equation}
S_n\rightarrow \frac{1}{2\pi i}\oint du \,\left(\frac{n}{n-1}\frac{1}{1-u}+\frac{\log u}{(u-1)^2}+{\cal O} (n-1)\right)\,\textrm{log}Z[u]\,.\label{chuta1}
\end{equation} 
 At $u=1$, $\textrm{log}Z[u]$ vanishes. We also have the identity $Z[e^{i 2 \pi a}]=Z[e^{-i 2 \pi a}]$, due to the Euclidean time reflection symmetry. This leads to
\begin{equation}
Z[u]=Z[1/u] \label{propi}\,.
\end{equation}
Then,  $\textrm{log}Z[u]$ should be at least of order $(u-1)^2$ near $u=1$. In consequence, the first term within the brackets does not have singularities inside the contour and does not contribute. Thus, the only singularity is the cut of the logarithm and we obtain
\begin{equation}
S(V)=-2\int_1^{\infty}d\lambda\,\, \frac{\log Z[-\lambda]}{(\lambda+1)^2}\,,\label{boti}
\end{equation} 
where we have used the property (\ref{propi}).

The fermion case can be treated similarly. The sum of residues in this case is over the roots of $(-1)^{n+1}$, giving
 \begin{equation}
S_n=\frac{1}{2\pi i (1-n)}\sum_{k=-(n-1)/2}^{(n-1)/2}\oint du \, \frac{\textrm{log}Z[u]}{u-e^{i 2 \pi k/n}}=\frac{1}{2\pi i (1-n)}\oint dv \,\frac{n\, v^{n-1}}{v^n+1}\,\textrm{log}Z[-v]\,.\label{chuta2}
\end{equation}
For the last integral in $v=-u$ we choose the contour in figure \ref{figu22} (the same as for scalars). Again, using property (\ref{propi}), the limit $n\rightarrow 1$ gives     
\begin{equation}
S(V)=2\int_1^{\infty}d\lambda\,\, \frac{\log Z[\lambda]}{(\lambda-1)^2}\,.\label{vein}
\end{equation}
where we assume $\left|\textrm{log} Z[\lambda]\right|/\lambda^x\rightarrow 0$ as $\lambda\rightarrow\infty$ for some $x<1$. 
In concrete examples, the result for $\log Z[e^{i 2 \pi a}]$ automatically gives a formula for the analytic extension $\log Z[u]$. Then the entropy is calculated by (\ref{boti}) or (\ref{vein}). 

\subsubsection{Methods for calculating the partition function}
One of the most powerful methods for computing the partition function of quadratic actions is the heat kernel method. To introduce it, consider a free scalar, and note that
\begin{equation}
W=-\log(Z)=\frac{1}{2} \log \det (m^2-\nabla ^2)=\frac{1}{2}\textrm{tr}\log(m^2-\nabla ^2)\,.\label{with}
\end{equation}
The partition function then reduces to an spectral function of a differential operator. 
The heat kernel is defined as $K(x,y,t)=\left<x\right|e^{ t \nabla^2}\left|y\right>$, and its trace writes
\begin{equation}
\zeta (t)=\textrm{tr}\, e^{ t \nabla^2} =\int dx \, K(x,x,t)\,.
\end{equation}
The free energy $W$ can be written in terms of this spectral function as
\begin{equation}
W=-\frac{1}{2}\int^\infty_\epsilon dt\, e^{-m^2 t} \frac{1}{t}\zeta (t)\,,
\end{equation}
where $\epsilon$ is a cutoff. 
The advantage of this equation if that, in contrast with (\ref{with}), it is written as a function of the trace of an operator which satisfies a local heat equation
\begin{equation}
\frac{\partial K}{\partial t}=\nabla^2 K\,, \hspace{1cm} K(x,y,0)=\delta(x-y)\,.
\end{equation}
This means that systematic expansions for short time $t$ of the trace of the heat kernel can be developed, leading to an expansion of the form (for a review see \cite{vasi})
\begin{equation}
\zeta (t)=\sum_{k\ge 0} t^{(k-D)/2} a_k\,.\label{ggg}
\end{equation}
Here $D=d+1$ is the manifold dimension, and $d$ is the dimension of space. In general, the heat kernel coefficients $a_k$ are integrals of local quantities depending on the different background tensors. This may include a smooth metric, interactions with external sources or fields, or a smooth boundary and boundary conditions. An expansion is known for different spins up to the first few $k$ in the general case.  

However, the application of the heat kernel expansions to the entanglement entropy in flat space vacuum is made difficult because of two reasons.  
The first one is that the short $t$ expansion is an ultraviolet one. Thus, the non divergent contributions, which represent genuine physical terms in the entanglement entropy, are the ones proportional to the coefficients $a_k$ with $k\ge d+1$. The case $k=d+1$ gives a logarithmically divergent contribution to the entanglement entropy, with universal coefficient which is proportional to the conformal anomaly (see section 3.3.3). Obtaining  these large-index coefficients becomes increasingly difficult for higher dimensions. 

The second reason is that the manifold has conical singularities along the boundary of the region $V$, and the standard expansions cannot be applied directly to this case. Expansions in presence of conical singularities in the limit of small deficit angle, with bulk curvature, and smooth $\partial V$ having zero extrinsic curvature, have been developed in \cite{dos} (see also \cite{dt}). They where applied to obtain logarithmic corrections of the black hole entropy in different scenarios  \cite{dos,cases}. Recently, the contribution of the extrinsic curvature was found in four dimensions, and used to calculate the logarithmic term of the vacuum entanglement entropy  in Minkowski space with smooth $\partial V$ \cite{una}. We will review this result in section 3.3.3.     

Contributions from a non smooth boundary become intractable with the heat kernel method. We will encounter this type of term in section 3.2. 

Another possible strategy to find the partition function consists in studying the associated Green function $G=(-\nabla^2+m^2)^{-1}$ on the manifold relevant to the problem. These are related by the identity
\begin{equation}
\frac{d}{dm^2}\log Z= -\frac{1}{2}\,\textrm{tr}  \, G\,.
\end{equation} 
 However, there are no general methods to calculate $G$ for this type of manifolds with a codimension one cut on a finite region, and a case by case analysis seems unavoidable. In some of the examples we review in section 3, we have used a method originally introduced by Myers in \cite{Myers} to deal with Neumann and Dirichlet boundary conditions on a finite cut in two dimensions. It essentially consists in exploiting the symmetries of the Helmholtz equation, even in the presence of the boundary conditions, by analyzing the singular behavior of $G(\vec r,\vec{r^{\prime}})$ at the end points of the cut, which effectively control the solutions of the source free equation. With this tool we could calculate the entropy for a single interval for massive free fields (sections 3.1.1 and 3.1.2) and the coefficient of the term logarithmic in the cutoff for a plane angular sector in two dimensions (section 3.2.1 and 3.2.2). 

In $d=1$ other techniques are also available. In particular the bosonization technique can be used to evaluate $Z(n)$ for the Dirac field (section 3.1.5). One expresses the fermionic current in terms of a dual scalar field $\phi$ as $j_{k}^{\mu }\to \frac{1}{\sqrt{\pi }}\epsilon ^{\mu \nu }\partial_{\nu}\phi$.
The resulting dual theory is the Sine Gordon theory, and the functional $Z(n)$ is given in terms of a sum of  correlators of local exponential operators \cite{fermion}. 
This is a particular case of the general one in one dimensional space, since $Z(n)$ can always be expressed as a correlator of twisting operators \cite{ccd} (see also \cite{cg}). These are non local relative to the ordinary fields, and effectively impose the boundary conditions. The correlators of twisting operators for general integrable massive quantum field theories have been studied  from a S matrix approach. In particular cases, an expansion of the correlators in form factors is known, and they can be evaluated as a sum over intermediate multi-particle states. These expansions have been exploited in relation with entanglement entropy in \cite{dc,ccd,d1}.

\subsection{Real time approach}
 In the real time approach one aims to compute directly the reduced density matrix corresponding to the global vacuum state in terms of correlators. One starts with a discrete version of the quantum field theory and eventually takes the continuum limit. 
The first calculations of the entanglement entropy where made in this way \cite{bombelli}, several years before the Euclidean approach was developed. This method has been mainly applied to numerical calculations in the lattice. Even if it has not been exploited as much as the Euclidean approach for analytic calculations, we think it is better suited to several generalizations. For example, it can be used in  calculations involving spatial sets which are not contained in a single spatial hyperplane in Minkowski space, for which the Euclidean approach becomes inapplicable. The interactions can be included in a straightforward way, at least at the perturbative level \cite{v1}. Also, the case of some states different from the vacuum seems to be tractable with this method. However, these applications have not been fully developed in the literature yet.
 
First, let us describe briefly how this approach was formulated originally.   For a system of harmonic oscillators with  Hamiltonian
\begin{equation}
H=\frac{1}{2}\sum_{i=1}^N \dot{\phi}_i^2 +\frac{1}{2}\sum_{i,j=1}^N \phi_i K_{ij}\, \phi_j\,\label{hache}
\end{equation}
the ground state wave function can be easily obtained by diagonalization
\begin{equation}
\Psi_0(\phi)=\left(\textrm{det}\frac{W}{\pi}\right)^{\frac{1}{4}}e^{-\frac{1}{2}\phi^T W \phi}\,,
\end{equation}
where $W=\sqrt{K}$.
 The corresponding density matrix in coordinate representation is  
\begin{equation}
\rho(\phi,\phi^\prime)=\Psi_0^*(\phi)\Psi_0(\phi^\prime)=\sqrt{\textrm{det}\frac{W}{\pi}}e^{-\frac{1}{2}\phi^T W \phi} e^{-\frac{1}{2}\phi^{\prime T} W \phi^\prime}\,.\label{refe}
\end{equation}

Then, one selects the variables $\phi^{(i)}$ and $\phi^{(o)}$ corresponding to the degrees of freedom inside and outside $V$. The trace over the $\phi^{(o)}$ can be done by integration of the Gaussian  in (\ref{refe}). Writing
\begin{equation}
W=\left(
\begin{array}{cc}
W_{(i)} & W_{(io)} \\
W_{(oi)} & W_{(o)}
\end{array} 
\right)\,,
\end{equation}   
in the base where the inside variables $q^{(i)}$ are the first ones, one arrives at
\begin{equation}
\rho_V(x,x^\prime)=\sqrt{\textrm{det}\frac{1-\Lambda}{\pi}}e^{-\frac{1}{2}x^T x} e^{-\frac{1}{2}x^{\prime T} x^\prime} e^{\frac{1}{4} (x+x^\prime)^T \Lambda (x+x^\prime)}\,,
\end{equation}
where  
\begin{eqnarray}
x &=&W_{(i)}^{1/2} \phi^{(i)}\,,\\
\Lambda &=& W_{(i)}^{-1/2} W_{(io)} W_{(o)}^{-1} W_{(oi)}  W_{(i)}^{-1/2}   \,.\label{baba}
\end{eqnarray}
The entropy follows by writing the density matrix as one for non coupled degrees of freedom by making a linear transformation of coordinates. Finally we have
\begin{equation}
S(V)=\textrm{tr} \left(\log (\frac{1-\Lambda/2+\sqrt{1-\Lambda}}{1-\Lambda+\sqrt{1-\Lambda}})-\frac{\Lambda\, \log \left(\frac{\Lambda}{2-\Lambda+2\sqrt{1-\Lambda}}\right)}{2 (1-\Lambda+\sqrt{1-\Lambda})}\right)\,.\label{la}
\end{equation}

This calculation has first been done in \cite{bombelli}. It was later rediscovered in \cite{srednicki} where it was applied to free quantum fields discretized in the radial direction in polar coordinates. For fermions a similar calculation, has been done in \cite{vidal}. 

In the following, we use a method which starts with a different perspective, and gives equivalent results in a way more suitable to generalizations. By its very definition, the reduced density matrix $\rho_V$ corresponding to the region $V$, is the state acting on the local algebra of operators in $V$ which leads to the same expectation values as the global vacuum state,
\begin{equation}
\left\langle O_{V}\right\rangle =\textrm{tr}(\rho
_{V}O_{V})\,,\label{nueve}
\end{equation}
 for any
operator $O_{V}$ localized inside a $V$. If we take the point of view of the Wightman theorem \cite{wightman}, where one defines a quantum field theory in terms of the correlators, this equation suggests that the knowledge of all the correlators inside $V$ must be enough to determine the density matrix $\rho_V$. This is specially simple in the free case, where the Wick theorem means that all the correlators are reduced to the two point function. This fact was
used by Peschel to give an expression for $\rho _{V}$ in
terms of correlators for free Boson and Fermion discrete systems \cite{peschel} (see also \cite{vidal}). Similar expressions for the local density matrices in terms of correlators where obtained long time ago by Araki \cite{araki}. His work is framed in a more general and mathematically oriented context, classifying the possible states satisfying the Wick theorem (called quasifree states) for algebras obeying the canonical commutation and anti-commutation relations.   

\subsubsection{Bosons}

The local Hermitian variables  $\phi _{i}$ and $\pi _{j}$ (coordinate and conjugate momentum) obey the canonical commutation relations 
\begin{equation}
[\phi _{i},\pi _{j}]=i\delta _{ij}\,,\hspace{1cm} [\phi _{i},\phi _{j}]=[\pi _{i},\pi _{j}]=0\,.
\end{equation} 
Take the two point correlators inside $V$ to be 
\begin{eqnarray}
\left\langle \phi _{i}\phi _{j}\right\rangle &=& X_{ij} \,, \hspace{1cm}\left\langle \pi _{i}\pi _{j}\right\rangle =P_{ij} \,, \label{atre}\\
 \left\langle \phi _{i}\pi _{j}\right\rangle &=& \left\langle \pi _{j}\phi _{i}\right\rangle^*=\frac{i}{2}\delta _{ij}\,.\label{need}
\end{eqnarray}
This last equation can be generalized to have $\left\langle \phi _{i}\pi _{j}\right\rangle + \left\langle \pi _{j}\phi _{i}\right\rangle\neq 0$, but (\ref{need}) is all we need for the vacuum state. 
The equations in (\ref{atre}) imply the matrices $X$ and $P$ are real Hermitian and positive. The positivity of $\left\langle (\phi _{l}+i \lambda_{lk} \pi_k )(\phi _{m}-i \lambda^*_{ms} \pi_s)\right\rangle $ for arbitrary constants $\lambda_{lk}$ implies that
\begin{equation}
X.P\ge \frac{1}{4}\label{popo}\,,
\end{equation} 
in matrix sense, what means that the eigenvalues of $X.P$ are greater than $1/4$. 

We assume all other non zero correlators are obtained from these by the Wick's   theorem 
\begin{equation}
\left<{\cal O} f_{i_1} f_{i_2}...f_{i_{2k}}\right>=\frac{1}{2^k k!}\sum_{\sigma}\left<{\cal O}f_{i_{\sigma(1)}} f_{i_{\sigma(2)}}\right>... \left<{\cal O} f_{i_{\sigma(2k-1)}} f_{i_{\sigma(2k)}}\right>\,,\label{quin}
\end{equation}
where the sum is over all the index permutations $\sigma$, the $f_i$ can be any of the field or momentum variables, and ${\cal O}$ is an ordering prescription, for example, ordering the products inside the expectation values with the field variables at the left and the momentum variables on the right. Once this equation holds for a specific ordering automatically holds for any other ordering.    
  
Let us consider now general independent creation and annihilation operators $a_{l}$, $a_l^\dagger$, 
with $
[a_{i},a_{j}^{\dagger }]=\delta _{ij}$, 
 which are 
expressed as linear combinations of the $\phi _{i}$ and $\pi _{j}$, $i$, $j\in V$,
\begin{eqnarray}
\phi_i&=&\alpha_{ij}^* a^\dagger_j+ \alpha_{ij} a_j\,,\label{sie}\\
\pi_i&=& -i \beta_{ij}^*  a^\dagger_j+ i \beta_{ij}  a_j\,.\label{oto} 
\end{eqnarray} 
The commutation relations between coordinates and momentum give 
\begin{equation}
\alpha^* \beta^T +\alpha \beta^\dagger =-1\,.
\end{equation}

At this point we note that an ansatz for the density matrix of the form \cite{peschel,chung}
\begin{equation}
\rho _{V}=K e^{-{\cal H}}=K\,e^{-\Sigma \epsilon _{l}a_{l}^{\dagger }a_{l}}\,,  \label{osc1}
\end{equation}
where  the normalization constant $K=\Pi_l (1-e^{-\epsilon_l} )$, leads automatically to the Wick property for the correlators  
$\left<{\cal O} f_{i_1} f_{i_2}...f_{i_{2k}}\right>=\textrm{tr}(\rho _{V}  {\cal O} f_{i_1} f_{i_2}...f_{i_{2k}})$ \cite{wick}. Here ${\cal H}$ is called the modular Hamiltonian corresponding to the region $V$ \cite{haag}, and (\ref{osc1}) proposes it quadratic in terms of the creation and annihilation operators. 
 The Wick property holds independently for every mode, as can be shown by direct computation,
\begin{equation}
\textrm{tr} \left[(1-e^{-\epsilon_i} ) e^{-\epsilon_i a^\dagger_i a_i}      a_i^{\dagger n}   a_i^n   \right]=n!\,\textrm{tr}\left[(1-e^{-\epsilon_i} ) e^{-\epsilon_i a^\dagger_i a_i}  a_i^{\dagger} a_i   \right]^n= n! (e^{\epsilon_i} -1)^{-n}\,.
\end{equation}
From linearity, the eq. (\ref{quin}) follows.    
Then, with this expression for $\rho_V$ we calculate $\textrm{tr} (\rho_V \phi_i\pi_j)=(i/2)\delta_{ij}$, $\textrm{tr} (\rho_V \phi_i\phi_j)=X_{ij}$, and $\textrm{tr} (\rho_V \pi_i\pi_j)= P_{ij}$. This gives  
\begin{eqnarray}
\alpha^* n \beta^T -\alpha (n+1)\beta^\dagger=\frac{1}{2}\,,\\
\alpha^* n \alpha^T+ \alpha (n+1) \alpha^\dagger=X \,,\\
\beta^* n \beta^T+ \beta (n+1) \beta^\dagger=P\,,
\end{eqnarray}
where $n$ is the diagonal matrix of the expectation value of the occupation number 
\begin{equation}
n_{kk}=\left<a^\dagger_k a_k\right>=(e^{\epsilon_k}-1)^{-1}\,.
\end{equation} 
These equations give that  
$\alpha =\alpha_1 \, U $ and $
\beta = \beta_1 \, U $,
with $U$ unitary and diagonal, and $\beta_1$ and $\alpha_1$ real. The matrix of phases $U$ can be reabsorbed in the definition of the $a_i$, so we set $U=1$. Then we have 
 $
\alpha=-\frac{1}{2} (\beta^T)^{-1}$, and 
\begin{equation}
\alpha \frac{1}{4}(2n+1)^2 \alpha^{-1}=X P\,.
\end{equation}
This last equation gives the spectra of the density matrix in terms of the spectrum of $X P$,  
\begin{equation}
(1/2)\coth(\epsilon_k/2)=\nu_k \,,
\end{equation}
 where $\nu_k$ are the (positive) eigenvalues of $C=\sqrt{XP}$. This means that, in the bosonic case, the modular Hamiltonian is definite positive.
 
 Inverting the relations (\ref{sie}) and (\ref{oto}) and replacing in (\ref{osc1}) we can write the density matrix as
\begin{equation}
\rho _{V}=K\,e^{-\sum_V \left( M_{ij} \phi_i \phi_j + N_{ij} \pi_i \pi_j \right)}\,,  \label{osc}
\end{equation}
where
\begin{eqnarray}
M&=&\frac{1}{4} \alpha^{-1\,T} \,\epsilon \,\alpha^{-1}=P\,\frac{1}{2 C}\log\left(\frac{C+\frac{1}{2}}{C-\frac{1}{2}}\right)\,,\\
N&=&\alpha \,\epsilon \,\alpha^T=\frac{1}{2 C}\log\left(\frac{C+\frac{1}{2}}{C-\frac{1}{2}}\right)\,X\,,
\end{eqnarray}
with $\epsilon$ the diagonal matrix of the $\epsilon_k$. The entropy is given by
\begin{equation}
S=\sum_l \left(-\log (1-e^{-\epsilon _{l}})+\frac{\epsilon _{l}\, e^{-\epsilon _{l}}}{
1-e^{-\epsilon _{l}}}\right)=\textrm{tr}\left(( C+1/2)\log (C+1/2)-(C-1/2)\log (C-1/2)\right)\,,  \label{for}
\end{equation}
which is positive thanks to  $C >1/2$, eq. (\ref{popo}). We also have
\begin{equation} 
\log (\textrm{tr}\rho^n)=-\textrm{tr}\left[ \log\left((C+1/2)^n-(C-1/2)^n\right)\right]\,.
\end{equation}

For a quadratic Hamiltonian for bosonic degrees of freedom with the form (\ref{hache}),   
$
H=\frac{1}{2}\sum \pi _{i}^{2}+\frac{1}{2}\sum_{ij}\phi _{i}K_{ij}\phi
_{j}$, the vacuum (ground state) correlators are given by 
\begin{eqnarray}
X_{ij} &=&\left\langle \phi _{i}\phi _{j}\right\rangle =\frac{1}{2}(K^{-
\frac{1}{2}})_{ij}\,,  \label{x} \\
P_{ij} &=&\left\langle \pi _{i}\pi _{j}\right\rangle =\frac{1}{2}(K^{\frac{1
}{2}})_{ij}\,. \label{p}
\end{eqnarray}
For the global state we have $X.P=1/4$, which has zero entropy, corresponding to a pure state. 

A straightforward calculation shows that in this case $C=\sqrt{XP}=1/2(1-\Lambda)^{-1/2}$ (where $\Lambda$ is given by (\ref{baba})), and the expression (\ref{for}) for the entropy is equivalent to (\ref{la}) obtained with the Hamiltonian approach. However, this formulation stresses the fact that we need only the correlators inside $V$, which is not apparent in the eq. (\ref{baba}), involving matrix elements in the exterior region. In some situations, the field correlators are known, and this knowledge may lead to a great reduction in computing time in numerical simulations, avoiding the use of an infrared cutoff. Besides, it makes more transparent the range of applicability of the formalism, which extends to all states satisfying the Wick theorem (Gaussian states). This includes states in thermal equilibrium and with chemical potential, or some vacuum states in curved space-time. 
 
In the continuum limit $C$ converges to an integral kernel. However, in this limit is not obvious how this depends on the correlators $\langle \phi(0) \phi(x)\rangle$ and $\langle \pi(0) \pi(x)\rangle$, since these do not define good operator kernels.    

\subsubsection{Fermions}
The local creation and annihilation operators $\psi_{i}^{\dagger }$, $\psi_{j}$
satisfy the anticommutation relations $
\{\psi_{i},\psi_{j}^{\dagger }\}=\delta
_{ij}$. 
 Let the two point correlators be
\begin{eqnarray}
\left< \psi_{i} \psi_{j}^\dagger
 \right> &=& C_{ij} \,,  \hspace{1cm}
\left< \psi_i^\dagger \psi_j \right> = \delta_{ij}- C_{ji} \,,\label{ffff}\\
\left< \psi_{i} \psi_{j}\right> &=& \left< \psi_{i}^\dagger \psi_{j}^\dagger\right>=0\,.\label{qv}
\end{eqnarray}
We also assume the Wick theorem holds and all non-zero multipoint correlators are obtained from the two point functions in the standard way
\begin{equation}
\left<\psi_{i_1}...\psi_{i_k}\psi^\dagger_{j_1}...\psi^\dagger_{j_k}\right>= (-1)^{\frac{k(k-1)}{2}}\sum_\sigma \epsilon_\sigma   \prod_{q=1}^k  \left< \psi_{i_q} \psi_{j_{\sigma(q)}}^\dagger
\right>\,,
\end{equation}
where the sum is over all permutations of the set of indices $j_1,..., j_k$, and $\epsilon_\sigma$ is the permutation signature. 
 The case when the correlators of Eq. (\ref{qv}) are different from zero can also be treated, but we will not need this generality here.   
The hermitian matrix $C_{ij}$ of correlators inside the region $V$ must determine completely the local state $\rho _{V}$. 
From (\ref{ffff}) $C_{ij}$ and $\delta_{ij}- C_{ji}$ are positive, and thus have eigenvalues in the interval $[0,1]$. For $V$ the total space, and when the global state is pure, $C$ is a projector, having eigenvalues $0$ or $1$. 
  
Analogously to the bosonic case, the correlators computed with the help of the reduced density matrix $\rho_V$ and eq. (\ref{nueve}) satisfy the Wick property if we have \cite{wick} 
  \begin{equation}
\rho _{V}=K e^{-{\cal H}}=K e^{-\sum_V H_{ij}\, \psi_{i}^\dagger\,. \psi_{j}}\label{corre}
\end{equation}
 Since $\rho_V$ must be Hermitian $H$, is also Hermitian. We can then diagonalize the exponent by the Bogoliuvov transformation
 $
d_{l}=U_{lm} \psi_m
$, 
with $U$ unitary in order to maintain the anticommutation relation, $\{d_i,d_j^\dagger\}=\delta_{ij}$.
One then chooses $U$ such that $U.H.U^\dagger=\{\epsilon_i \}$ is a diagonal matrix where the $\epsilon_i$ are the eigenvalues of $H$. One has
\begin{equation}
\rho _{V}=\prod \frac{e^{-\epsilon _{l}d_{l}^{\dagger }d_{l}}
}{\left( 1+e^{-\epsilon _{l}}\right) }, \label{diago}
\end{equation}
and from here we specify the constant $K=\textrm{det}(1+e^{-H})^{-1}$.

The relation between $H$ and $C$ follows from
\begin{equation}
K\,\, \textrm{tr}\, \left(e^{-\sum_V H_{lm}\, \psi_{l}^\dagger \psi_{m}} \psi_i \psi_j^\dagger \right)= C_{ij}\,.
\end{equation}
Using the diagonalization (\ref{diago}) one obtains that the eigenvalues 
$\epsilon _{l}$ of $H$ and the eigenvalues $\nu _{l}$ of $C$ are related by 
\begin{equation}
e^{\epsilon _{l}}=\frac{\nu _{l}}{1-\nu _{l}}\,,
\end{equation}
or, in matrix notation,
\begin{equation}
H=-\log \left( C^{\,-1}-1\right)\,.\label{cincuenta}
\end{equation}
With $\nu_l \in (0,1)$ the one particle Hamiltonian ${\cal H}$ is  not defined positive. However,  the large negative energies do not contribute much to the entropy, since the corresponding degrees of freedom become saturated at occupation number $1$ (the equivalent to the Dirac sea). The large positive energies saturate at occupation number $0$. 

The entropy and $\log \textrm{tr}\rho ^{n}\,$ can be evaluated as a sum over independent modes
\begin{eqnarray}
 S(V)=\Sigma_l \left( \log (1+e^{-\epsilon _{l}})+\frac{\epsilon _{l} e^{-\epsilon _{l}}
}{1+e^{-\epsilon _{l}}}\right) =-\textrm{tr}\left((1-C)\,\log (1-C)+C \,\log (C) \right)\,, \label{tiuno}\\
\log \textrm{tr}\rho_V^{n} =\Sigma_l\left(\log (1+e^{-n\epsilon _{l}})-n\log (1+e^{-\epsilon
_{l}})\right) =\textrm{tr} \left( \log ((1-C)^{n}+C^{n})\right)\,.\hspace{1.2cm}\label{44}
\end{eqnarray}

The typical case where the Wick theorem holds is for a quadratic global Hamiltonian of the form 
\begin{equation}
{\cal H}=\sum_{i,j}M_{ij}\psi_{i}^{\dagger }\psi_{j}\,,  \label{hami}
\end{equation}
choosing any Fermi level and temperature, since in this case the  global density matrix is an exponential of a quadratic form in the creation and annihilation operators.

In quantum field theory applications one takes the vacuum (half filled) state of a Hamiltonian with symmetric spectrum around the origin.
 The explicit form of the correlator in this case is a projector
\begin{equation}
C=\theta (-M)\,,
\end{equation}
where $\,\theta (x)$ is the step function. For the whole space we then have zero entropy. The reduction to a region makes the eigenvalues of $C$ lie  between zero and one. In this case we also have a total symmetry between $\psi_i$ and $\psi^\dagger_i$, (a local charge conjugation symmetry) which leads to the same spectrum for $C$ and $1-C$. Therefore the spectrum of energies of the modular Hamiltonian is also symmetric around $0$.

\noindent \textbf{Density matrix for a Dirac field}
 
\noindent As an application of these formulas to quantum field theory, let us consider a free Dirac field in $d+1$ dimensions (an early treatment of the continuum limit for fermions can be found in \cite{p1}). The field satisfies the Dirac equation $
(i\partial_\mu \gamma^\mu -m) \, \Psi =0$, 
 with $\gamma^\mu$ the Dirac matrices, and the 
canonical (equal-time) anticommutation relations
\begin{equation}
\left. \left\{\Psi_i(\vec{x},t),\Psi^{\dagger}_j(\vec{y},t^\prime)\right\}\right|_{t=t^\prime}=\delta^d(\vec{x}-\vec{y})\delta_{ij}\,.
\end{equation}
We can discretize the field algebra (at $t=0$) by using the smoothed field operators
\begin{equation}
\psi_{n}=\int_V dx^d\, \Psi_i(\vec{x},0) \alpha^i_n (\vec{x})\,,\label{rela}
\end{equation}  
where $\alpha_n(\vec{x})$ is an orthonormal base of spinor functions on $V$, $
\int_V dx^d\, \alpha_n^\dagger (\vec{x}) \alpha_m (\vec{x})=\delta_{n,m}$, 
and the discrete field operators satisfy $
\{\psi_m,\psi_n^\dagger\}=\delta_{m,n}$.
The results of the previous section can be directly applied to these set of discrete operators. After that one can recover the field language with the help of (\ref{rela}). 

Let then the set $V$ be an arbitrary region contained in a surface of constant time. We have the expressions analogous to (\ref{corre}) and (\ref{cincuenta}) 
\begin{eqnarray}
\rho &=& K\, e^{-{\cal H}}\,, \\
{\cal H}&=&\int_V dx^d\,dy^d \,\Psi_i^{\dagger}(\vec{x},0)H_{ij}(\vec{x},\vec{y}) \Psi_j(\vec{y},0)\,,\label{rho2}\\
H&=&-\log (C^{-1}-1 )\,,
\end{eqnarray}
where the field correlator is 
\begin{equation}
C(\vec{x},\vec{y})=\left. \left\langle0 \right|\Psi(\vec{x},0)\,\Psi^\dagger(\vec{y},0) \left|0\right\rangle \right|_{\vec{x},\vec{y}\in V}=\int \frac{dp^d}{(2\pi)^d}  \frac{(p_\mu \gamma^\mu +m)}{2 \sqrt{p^2+m^2}}\gamma^0 e^{-i \vec{p}(\vec{x}-\vec{y})}\,.
\label{kernel}
\end{equation}

The expression for the entropy is again (\ref{tiuno}), but now $C$ is an operator with kernel (\ref{kernel}) rather than a matrix, and (\ref{tiuno}) requires a regularization. 
 The more general case of an arbitrary spatial region not contained in a single spatial hyperplane is treated in \cite{futuro}.

\subsection{Direct relation between the Euclidean and the real time approaches}
The expression for the entropy in the Euclidean approach can be directly transformed into the real time approach formulas (\ref{for}) and (\ref{tiuno}) by evaluation of the functional integral $ Z[\lambda]$ in terms of boundary operators. A similar treatment has been applied to the Casimir effect in \cite{fos}.

Let us start with the Dirac field. The boundary condition in $Z[\lambda]$ , with $\lambda=e^{i 2 \pi a}$, is taken into account with the modified action $S[\bar{\Psi},\Psi]=S_0[\bar{\Psi},\Psi]+\delta S_V[\bar{\Psi},\Psi]$, where 
\begin{eqnarray}
S_0[\bar{\Psi},\Psi]&=&\int dx^{d+1}\,\bar{\Psi} (\gamma^\mu \partial_\mu + m) \Psi \,,\\
\delta S_V&=& \int dx^{d+1}\, (1-\lambda) \delta(\tau)\chi_V(\vec{x}) \bar{\Psi}(x) \gamma^0\Psi(x)= (1-\lambda) \int_V dx^d\, \bar{\Psi}\gamma^0 \Psi \,,
\label{86}
\end{eqnarray}
where $\tau$ is the Euclidean time coordinate,  $\chi_V(\vec{x})$ equal to one for $\vec{x}\in V$ and $0$ outside, and the last integral over the spatial $d$ dimensional set $V$.  This is because this new action leads to the same classical solutions and boundary conditions as the original problem. It is also possible to understand the term $\delta S_V$ as produced by an external gauge field vanishing outside  $V$, which is pure gauge everywhere except at $\partial V$, and which has the effect of imposing the correct boundary conditions (see section 3.1.5).
Strictly speaking, in a discretization of the path integral, the term in (\ref{86}) has to be understood as proportional to the product $\bar{\Psi}^+(x) \gamma^0\Psi^-(x)$ of the fields located on different sides of the cut.

Writing the partition function in terms of an auxiliary Grassmann field living in $V$ in order to linearize $\delta S_V$ we have
\begin{equation}
 Z[\lambda]=\int {\cal D}\bar{\Psi}{\cal D}\Psi{\cal D}\bar{\xi} {\cal D} \xi\,\, e^{-S_0[\bar{\Psi},\Psi]+\int_V dx^d \, \bar{\Psi} \xi + \bar \xi  \Psi+ \frac{1}{1-\lambda}\bar{\xi} \gamma^0\xi}\,. \label{niu}
\end{equation}
Integrating first over $\Psi$, $\bar{\Psi}$ we have 
\begin{equation}
 Z[\lambda]=\int {\cal D}\bar{\xi} {\cal D} \xi\,\, e^{-\int_V dx^d \,\int_V dy^d \,\bar{\xi}(x) \langle \Psi(x) \bar{\Psi}(y)\rangle_E \xi(y) +  \frac{1}{1-\lambda}\int_V dx^d \bar{\xi} \gamma^0\xi}=\det\left(f_\lambda \left(1+(\lambda-1) C\right)\right), \label{pista}
\end{equation}
where, as in section 2.2.2, $C$ is the Minkowskian correlator $C(x,y)=\langle\Psi(x)\Psi^\dagger(y)\rangle=\langle\Psi(x)\bar{\Psi}(y)\gamma^0\rangle_E$ inside $V$, and $f_\lambda$ is an unimportant normalization factor which cannot change the universal terms in $\log Z[\lambda]$. The analytic extension of $Z[\lambda]$ for $|\lambda|\neq 1$ is done directly with formula (\ref{pista}). The symmetry $Z[\lambda]=Z[1/\lambda]$ (eq. (\ref{propi})) (for the universal terms) is mapped here  to the identity of the spectra of $C_V\in (0,1)$ and $1-C_V$.
We can choose $f_\lambda=\lambda^{-1/2}$ in order to have $Z[1]=1$, and $Z[\lambda]=Z[1/\lambda]$ exactly.  Using (\ref{pista}) in the formula (\ref{vein}) obtained from the analytic continuation of $S_n$ in section 2.1.3, and taking into account the spectral properties of $C$, we obtain the correct formula for the entropy (\ref{tiuno}).

The boundary condition for the scalar partition function can be similarly implemented by adding a term to the  first order action, 
\begin{equation}
S[\phi,\pi]=\int dx^{d+1} \left(\pi\dot{\phi}^*+\pi^*\dot{\phi}-\left(\pi \pi^*+\nabla\phi \nabla \phi^*+m^2 \phi \,\phi^*\right)\right)+\int_V dx^d ((1-\lambda) \phi\, \pi^*+(1-\lambda^{-1}) \phi^*\, \pi)\,.
\end{equation}    
Here $\lambda$ is again a phase factor.    
Writing the corresponding partition function in terms of auxiliary fields living on $V$, we get
 \begin{equation}
 Z[\lambda]=\int {\cal D}\pi^*{\cal D}\pi{\cal D}\phi^*{\cal D}\phi{\cal D}\xi_1^* {\cal D}\xi_1{\cal D}\xi_2^* {\cal D}\xi_2\,\, e^{-S_0[\pi,\phi]+\int_V dx^d\, (\pi^* \xi_1 +\xi_2^* \phi +\pi \xi_1^* +\xi_2 \phi^* +\frac{1}{1-\lambda} \xi_2^* \xi_1+\frac{1}{1-\lambda^{-1}} \xi_2 \xi_1^*)}\,.
 \label{nio}\end{equation}
We can now integrate over $\pi$ and $\phi$ first, giving
\begin{eqnarray}
 && Z[\lambda]=\int {\cal D}\xi_1^* {\cal D}\xi_1{\cal D}\xi_2^* {\cal D}\xi_2\,\, \exp\left(\int_V dx^d \, dy^d\,( \xi_1^*(x)\langle\pi(x)\pi^*(y) \rangle_E \xi_1(y)+\xi_2^*(x)\langle\phi(x)\phi^*(y) \rangle_E \xi_2(y)\right.  \label{twq}\\ &&+\left.(\xi_1^*(x)\langle\pi(x)\phi^*(y)\rangle_E\xi_2(y)+\textrm{h.c.})+\int_V dx^d \,(\frac{\xi_2^* \xi_1}{1-\lambda} +\frac{\xi_2 \xi_1^*}{1-\lambda^{-1}}) \right)= \textrm{det}^{-1}\left(f_\lambda\left(1- \frac{4(1-\lambda)^2}{(1+\lambda)^2}C^2\right)\right)\,,\nonumber
\end{eqnarray}
where $C=\sqrt{XP}$, with $X$ and $P$ the operators with kernel given respectively by the Minkowskian correlators $\langle\phi(x)\phi^*(y) \rangle=\langle\phi(x)\phi^*(y) \rangle_E$ and $\langle\pi(x)\pi^*(y) \rangle=-\langle\pi(x)\pi^*(y) \rangle_E$ restricted to $V$ (see section 2.2.1). We have also used $\langle\pi(x) \phi^*(y) \rangle_E=\langle\phi(x) \pi^*(y) \rangle_E=-1/2 \,\delta(x-y)$. When (\ref{twq}) is inserted in (\ref{boti}), it gives the correct formula for the entropy (\ref{for}), obtained with the real time formalism.  

It follows from (\ref{pista}) and (\ref{twq}) and the spectral properties of the operator $C$ in each case, that the universal terms in $\log Z[\lambda]$ (disregarding a global constant independent of $C$) satisfy the conditions for $|\lambda|\rightarrow \infty$ which were assumed in section 2.1.3. 

We note  that (\ref{niu}) and (\ref{nio}) give definitions for the partition functions $\log Z[\lambda]$ which extend analytically out of the range $|\lambda|=1$. The induced boundary conditions in this case are given by a factor $\lambda$ for the fields when crossing the cut, but a factor $\lambda^{-1}$ for the conjugate momentum.   

\section{Exact results}

\subsection{One spatial dimension}

\subsubsection{Single interval for a massive scalar field}
In this section we review the calculation of the universal part in $S_n$ and $S$, for a massive scalar field in a single interval $[L_2,L_1]$ presented in \cite{boson}. We use the Euclidean time method and
calculate the partition function $Z[\lambda]$ for a complex scalar from the Green function on a cut plane,
\begin{equation}
 \partial_{m^2}\log Z[\lambda]= -\int dr^d G_\lambda(\vec{r},\vec{r})\,.\label{gdgd}
\end{equation} 
 The singular behavior of the variations of the Green function near the end points of the cut under the action of the symmetries gives us the mechanism to find an exact expression for $\log Z$ in terms of the solution of
a second order non linear differential equation of the Painlev\'e V type. This method was first used in \cite{Myers}. 

Let us take the case $\lambda=e^{i 2 \pi a}$ with $a\in[0,1)$. The Green function $G(\vec{r},\vec{r^\prime})$ is uniquely defined by the following three requirements:

\noindent {\bf a.-} It satisfies the Helmholtz equation with a point like source
\begin{equation}
\left( -\Delta _{\vec{r}}+m^{2}\right) G(\vec{r},\vec{r}^{\prime})=\delta (
\vec{r}-\vec{r}^{\prime})\,. \label{g1}
\end{equation}

\noindent {\bf b.-} The boundary conditions are (they also hold for the Green function derivatives)
\begin{eqnarray}
\lim_{\epsilon \rightarrow 0^{+}}G((x,\epsilon ),\vec{r}^{^{\prime
}})&=&e^{i2\pi a }\lim_{\epsilon \rightarrow 0^{+}}G((x,-\epsilon ),\vec{r
}^{\prime}) \hspace{1cm} \textrm{for}\,\, x \in [L_{2},L_{1}]\,, \label{bcc1}\\
\lim_{\vert \vec{r} \vert \rightarrow \infty}G(\vec{r},\vec{r}^{\prime
})&=&0 \,.
\end{eqnarray}

\noindent {\bf c.-} $G(\vec{r},\vec{r^\prime})$ is bounded everywhere (including the cut) except at $\vec{r}=\vec{r}^{\prime}$.

\smallskip
\noindent This last requirement follows from the asymptotic condition (\ref{chia}).  
We will write the Green function as $G(z,z^{\prime})$ as a
shortcut of $G(z,\bar{z},z^{\prime},\bar{z}^{\prime},L_{1},L_{2})$,
where $z$ and $\bar{z}$ are the complex coordinates $x+iy$ and $x-iy$. 
It is Hermitian  $
G(z,z^{\prime})=G(z^{\prime},z)^{*}$, 
 and the time reflection symmetry gives $
G(z,z^{\prime})^{*}=G(\bar{z},\bar{z}^{\prime})$. 
 The
reflexion operation 
\begin{equation}
R\,\,(x,y)=(L_{1}+L_{2}-x,y) \label{refle}
\end{equation}
leaves the Helmholtz equation, the cut, and boundary conditions invariant. Thus we have 
\begin{equation}
G(z,z^{\prime})=G(Rz,Rz^{\prime})\,. \label{sime}
\end{equation}

Due to the boundary conditions, near the end points of the interval $[L_{2},L_{1}]$ the Green function must
have branch cut singularities. The requirement that
the function must remain bounded on the cut and the equation (\ref{g1})
imply that the most singular terms of $G(z,z^{\prime})$ for $z$ near $
L_{1}$ (and fixed $z^{\prime}$) have to be of the form
\begin{equation}
G(z,z^{\prime})\sim (z-L_{1})^{a }S_{1}(z^{\prime})+(\bar{z}
-L_{1})^{1-a }S_{2}(z^{\prime}) \,. \label{equ2}
\end{equation}
 Note that the contributions at this order must be  
analytic or anti-analytic in $z$ in
order to cancel the Laplacian term in (\ref{g1}).

In the following the fact that, due to the uniqueness of the solution, a function which satisfies the Helmholtz equation and is bounded everywhere including the cut, must vanish identically, is used repeatedly as a main argument in the calculation. This means that carefully analyzing the singular behavior at the extreme points of the cut of various quantities (formed by $G(z,z^\prime)$, $S_1(z)$, $S_2(z)$, or their derivatives) and combining them in order to cancel these singularities, one can actually construct equations which are valid everywhere.   
For example, from the analysis of the singular behavior of the derivatives $\partial_{L_1}G$ and $\partial_{L_2}G$ at the singular points, using (\ref{sime}) and (\ref{equ2}), we obtain the following fundamental relations
\begin{eqnarray}
\partial _{L_{1}}G(z,z^{\prime})&=&S(z)^{\dagger
}AS(z^{\prime})  \label{f1}\,,\\
\partial _{L_{2}}S(z)&=&\gamma \,S(Rz) \,, \label{a}
\end{eqnarray}
where $S$ is the vector with components $S_1$ and $S_2$, $\gamma $ is a certain (unknow up to this point) matrix function of $L=L_{1}-L_{2}$, and $A$ is a constant Hermitian matrix. The solution for the half infinity cut which can be studied with standard methods gives $A=-4\pi (1-a )a \,\sigma _{1}$, with $\sigma_1$  the Pauli matrix.
It also holds for consistency of these equations, 
\begin{equation}
\gamma ^{\dagger }=A\gamma A^{-1}\,. \label{129}
\end{equation}
Then, the equations (\ref{equ2}) and (\ref{f1}) lead to the singular behavior 
\begin{equation}
S(z)\sim \frac{1}{4\pi }\left(
\begin{array}{l}
\frac{1}{a }(z-L_{1})^{-a } \\
\frac{1}{(1-a )}(\bar{z}-L_{1})^{a -1}
\end{array}
\right)  \label{wr}
\end{equation}
for $z$ in the vicinity of $L_{1}$. 

The equations (\ref{gdgd}) and (\ref{f1}) allows us to express the partition function in terms of the functions $S_1$ and $S_2$
\begin{equation}
\partial _{L}\partial _{m^{2}}\log Z=-\int S^{\dagger
}A\;S=8\pi a \left( 1-a \right)H(L)\,,\label{partfunc}
\end{equation}
with
\begin{equation}
H(L) =\int dxdy\,S_{1}^{*}(z)S_{2}(z) \,. \label{hh}
\end{equation}
Then, in order to compute $\log Z$ we need more information on $S(z)$.

With this aim, we exploit the symmetries that the Helmholtz equation has without imposing boundary conditions on the cut, to find relations for $S$ and its derivatives. The idea is that due to translation and rotation symmetries, $\partial_y S$ and $\partial_\theta S$, where $\partial _{\theta }=x\partial
_{y}-y\partial _{x}$ is the rotation operator, also satisfy the Helmholtz equation and boundary conditions. Finding combinations which are free from divergences one finds the following equations 
 \begin{eqnarray}
 \partial_y S(z)&=& i\{\gamma,\sigma_3\} S(Rz)+i \sigma_3 \partial_x S(z)- \xi S(z)\,,\label{1111}\\
\partial _{\theta} S(z)&=&L_1\partial_{y} S(z)+iqS(z)-iL\gamma
\sigma _{3}S(Rz) \label{138} \,.
\end{eqnarray}
Here $\xi$ is another unknown matrix function of $L$ and 
\begin{equation}
q=\left(
\begin{array}{ll}
-a &  \\
& 1-a
\end{array}
\right)\,.
\end{equation}
The $R$ reflected equations to (\ref{1111}) and (\ref{138}) also hold. The consistency of these equations with the Helmholtz equation and (\ref{a}) gives the algebraic equations
\begin{eqnarray}
\xi^{\dagger}A+A\xi&=&0
\label{132}\,\\
\left\{ \xi ,\sigma _{3}\right\} &=&0 \,,\label{133}\\
\left\{ \left\{ \gamma ,\sigma _{3}\right\} ,\xi \right\} &=&0\,, \label{134}\\
\left( m^{2}+\left\{ \gamma ,\sigma _{3}\right\} ^{2}-\xi ^{2}\right) &=&0\,.\label{135}
\end{eqnarray}
and the differential equation
\begin{equation}
\xi =\frac{i}{L}\left( L\gamma ^{-1}\frac{d\gamma }{dL}\sigma _{3}+\gamma
^{-1}q\gamma +q+\sigma _{3}\right) \,. \label{eku}
\end{equation}

The algebraic equations (\ref{129}), (\ref{132} - \ref{135}) for the matrices are solved using the
parametrization
\begin{equation}
\gamma =\frac{m}{2}\left(
\begin{array}{ll}
u & b \\
c & u
\end{array}
\right)
\hspace{2cm};\hspace{2cm}\xi
=m\left(
\begin{array}{ll}
0 & i\beta _{1} \\
-i\beta _{2} & 0
\end{array}
\right)\,,
\end{equation}
where $u$, $b$, $c$, $\beta _{1}$, and $\beta _{2}$ are real functions of $t=mL$, and
$
u^{2}+1=\beta _{1}\beta _{2}$. 
From this and the differential equation (\ref{eku}) it follows that all variables can be expressed in terms of $u$ and $u^\prime$, and we have
\begin{eqnarray}
u^{\prime \prime } + \frac{1}{t}u^{\prime}-\frac{u}{1+u^{2}}
u^{\prime 2}-u(1+u^{2})-\frac{4u\left( a -\frac{1}{2}\right) ^{2}}{
t^{2}(1+u^{2})}=0 \label{ecdif}\,.
\end{eqnarray}
This nonlinear second order ordinary differential equation can be transformed to take 
 the form of a 
 Painlev\'e V equation \cite{fermion,ince}. 

In order to obtain a boundary condition for (\ref{ecdif}) consider the Green function $G_0(z,z^\prime)$ of the Helmholtz equation without the cut, and the equation
\begin{equation}
\partial_\mu \left(S_1(x)\partial_\mu G_0(x,x_1)-G_0(x,x_1)\partial_\mu S_1(x)\right)=-\delta^2(x-x_1) S_1(x)\,.
\end{equation} 
Integrating this equation on the plane we have
\begin{equation}
\oint dx\, \left(S_1(x)\partial_\mu G_0(x,x_1)-G_0(x,x_1)\partial_\mu S_1(x)\right)=-S_1(x_1)\,,\label{peron}
\end{equation}
 where the integration contour is around the cut $[L_2,L_1]$. Then one can use a massless limit expansion $G_0(0,r)\sim -1/(2\pi)(\log(r m/2)+\gamma_E)$, where $\gamma_E$ is the Euler constant, and the massless limit solution for $S_1(x)$ of eqs. (\ref{1111}-\ref{138}), which is given in terms of hypergeometric functions, in eq. (\ref{peron}), in order to derive the boundary condition 
\begin{equation}
u(t) \rightarrow  \frac{-1}{t \,(\log t+\kappa_S)}-a(a-1)t\,( \log(t)+\kappa_S)+...\;\;\;\;\;\textrm{as}\;\;t\to 0\,,\label{bbb1}
\end{equation}
with $\kappa_S=-\log(2)+2\gamma_E+\frac{\psi[a]+\psi[1-a]}{2}$ and $\psi[a]$ the digamma function. $u(t)$ admits a series expansion in terms of powers of $t$ and $\log(t)$ around the origin, with coefficients which are totally determined by $\kappa_S$, and the differential equation.

The long distance limit follows from the connection formulae for Painlev\'e equations \cite{rims}, or the form factor expansion (see section 3.1.6)
\begin{equation}
u_{a}(t)\rightarrow \frac{2}{\pi} \sin (a \pi) K_{1-2 a} (t)\,\,\,\,\,\,\,\, \textrm{as} \,\,\,\, t\rightarrow \infty \label{infinito}\,.
\end{equation}

\begin{figure} [tbp]
\centering
\leavevmode
\epsfxsize=8cm
\bigskip
\epsfbox{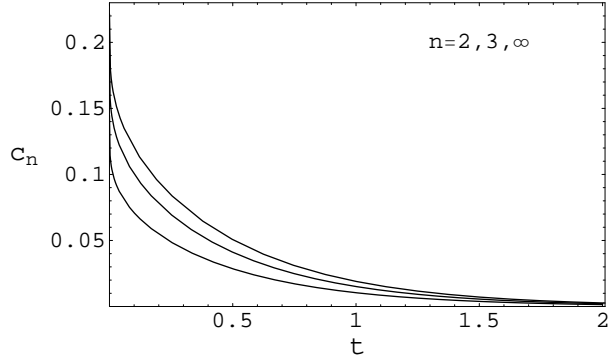}
\caption{The $c_n$ functions for a scalar field, with, from top to bottom, $n=2$, $3$, and $n\rightarrow \infty$. Here $t=m L$.} \label{fii}
\end{figure}

So far, we have solved partially the problem since we still do not have an explicit expression for (\ref{hh}).
To find this quantity, we define the following auxiliary real integrals 
\begin{eqnarray}
B_{1}(L) &=&\int dxdy\,S_{1}^{*}(z)S_{1}(Rz)\,, \\
B_{2}(L) &=&\int dxdy\,S_{2}^{*}(z)S_{2}(Rz) \,,\\
B_{12}(L) &=&\int dxdy\,S_{2}^{*}(z)S_{1}(Rz) \,,\\
X_{1}(L) &=&\int dxdy\,S_{1}^{*}(z)S_{1}(z) \,,\label{x1}\\
X_{2}(L) &=&\int dxdy\,S_{2}^{*}(z)S_{2}(z)\,.\label{x2}
\end{eqnarray}
Using eqs. (\ref{a}), (\ref{1111}), and (\ref{138}) on these expressions, a complete set of liner equations among these capital letter variables can be found. 
 As a result, all the capital letter 
variables can be written in terms of the functions $u$, $u^\prime$ by solving a linear system. In particular we get
\begin{equation}
H=\left(16 \pi a \left( 1-a \right) m\right)^{-1} tu^{2}\,.\label{h}
\end{equation}

Combining (\ref{partfunc}) and (\ref{h}) we have 
\begin{equation}
w_a=L\partial_L\log Z[e^{i 2\pi a}] =-\int_{t}^{\infty}dy\,\,y\,\,u^{2}(y)\,. \label{kuu}
\end{equation}

Therefore, the final expressions for the $c$ functions (\ref{centralcharge}) and (\ref{charge1}) are
\begin{eqnarray}
c_n(t)&=&\frac{1}{2(1-n)}\sum_{k=1}^{n-1} w_{k/n}(t)\label{suma}\,,\\
c(t)&=&-\frac{1}{2}\int_0^{\infty}db \frac{\pi}{\cosh(\pi b)^2}w_{-i b+1/2}(t)\label{ctint}\,,
\end{eqnarray}
Eq. (\ref{ctint}) follows from (\ref{boti}) by using $\lambda=e^{i 2\pi a}$, and $a=-i b+1/2$, where $b\in (0, \infty)$.
\begin{figure} [tbp]
\centering
\leavevmode
\epsfxsize=7.5cm
\bigskip
\epsfbox{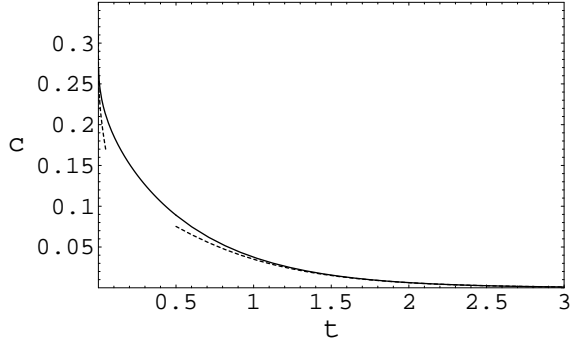}
\caption{The function $c(t)$ for a real scalar. At the origin it takes the value $1/3=0.33...$ and decays exponentially at infinity. The dashed curves are the leading short and long distance approximations.}\label{fiiii}
\end{figure}
We have introduced a $1/2$ factor in (\ref{suma}) and (\ref{ctint}) because we want to present the results for a real scalar  instead of a complex one. The function $c_n$ is plotted in figure \ref{fii} for $n=2$, $3$ and $n\rightarrow \infty$. The function $c(t)$ is shown in figure \ref{fiiii}.
The leading long distance terms on the entropy functions are
\begin{eqnarray}
c_{n }(t)&=&\frac{n}{n-1}\frac{e^{-2t}}{4\pi}+{\cal O}(\frac{e^{-2t}}{t})\,,\label{ccnn}\\
c(t)&\sim& \frac{1}{4}\;t \;K_1(2t) \label{cctt}\,.
\end{eqnarray}
Note from (\ref{ccnn}) that the limits $n\rightarrow1$ and $t\rightarrow\infty$ do not commute (see further details in \cite{fermion,dc}). Formula (\ref{cctt}) follows directly from (\ref{infinito}) and (\ref{ctint}). The short distance expansions read
\begin{eqnarray}
c_n(t)&=&\frac{1+n}{6 n}+\frac{1}{2\log(t)}+{\cal O}\left(\log^{-2}(t)\right)\,,\\
c(t)&=&\frac{1}{3}+\frac{1}{2\log(t)}+{\cal O}\left(\log^{-2}(t)\right)\,.\label{cuyas}
\end{eqnarray}
The constant term corresponds to the conformal case for which there is a general result identifying $c=C_V/3$, where $C_V$ is the Virasoro central charge \cite{ci,cc}. Here it is $C_V=1$ since we are considering a real scalar field.

The sharp cusp of $c(t)$ at the origin, due to the $1/\log(t)$ term in (\ref{cuyas}), is related to an infrared divergence for the entropy in the massless limit. In fact, for $m\rightarrow 0$ this behavior gives for the entropy differences
\begin{equation}
S(L)-S(L_0)=\int_{L_0}^L dl\, \frac{c(m l)}{l}\sim
\frac{1}{3}\log(L/L_0)+\frac{1}{2} \log\left(-\log(m\,L)\right)-\frac{1}{2} \log\left(-\log(m\,L_0)\right)\,.
\end{equation}    
 This suggests an infrared divergence $S(L)\sim 1/2 \log(-\log(m))$ for the entropy of any set. This fact can be checked by numerical simulations on the lattice, and has the following heuristic explanation. In the massless limit the homogeneous component of the field is a zero mode. The correlation function diverges logarithmically with the mass, and thus the typical size of the fluctuations on the homogeneous mode grows as $(-\log (m))^{1/2}$. Correspondingly, the entropy grows as the logarithm of this volume in field space \cite{unruh1}, and    
 becomes infrared divergent $S(L)\sim 1/2\log(-\log(m))$. This term in the entanglement entropy is  independent of the number of components of the set, due to its infrared origin. In consequence, the mutual information $I(A,B)\sim \frac{1}{2}\,\log(-\log(m))$ is also infrared divergent.

\subsubsection{Single interval for a massive Dirac field}
In this section we present a derivation of the entropic $c$ function for a massive Dirac field by relating the Dirac Green function to the scalar one studied in 3.1.1. This derivation is different from the one in \cite{fermion}, which is discussed later in 3.1.6.

The functional $Z[e^{i 2 \pi a}]$ can be calculated exploiting the relation between the free energy and the Green function  
\begin{equation}
\frac{d\log Z}{dm}=\textrm{tr} \, G_D\,,\label{pero}
\end{equation}
where $m$ is the field mass. 
 We will take advantage of what we learned from the scalar case in the previous section. The Euclidean Green function $G_D$ satisfies the equation
\begin{equation}
(\gamma_{\mu}\partial_{\mu}+m)G_D(z,z^{\prime})=\mathbb I \delta(x-x^{\prime})\delta(y-y^{\prime})\,. \label{deq}
\end{equation}
For definiteness we choose the Euclidean gamma matrices as $\gamma^1=\sigma^1$ and $\gamma^2=\sigma^3$.
For the scalar Green function $G_S(z,z^{\prime})$ we have
\begin{equation}
(\Delta-m^2)G_S(z,z^{\prime})=-\delta(x-x^{\prime})\delta(y-y^{\prime})\,.
\end{equation}
Then
\begin{equation}
(\Delta-m^2)(G_D(z,z^{\prime})-\mathbb I~m~G_S(z,z^{\prime}))=\begin{pmatrix}\delta(x-x^{\prime})\delta^{\prime}(y-y^{\prime})&\delta^{\prime}(x-x^{\prime})\delta(y-y^{\prime})\\\delta^{\prime}(x-x^{\prime})\delta(y-y^{\prime})&-\delta(x-x^{\prime})\delta^{\prime}(y-y^{\prime}) \end{pmatrix}\label{gsgd}\,,
\end{equation}
where we have used that
\begin{eqnarray}
(\Delta-m^2)G_D(z,z^{\prime})&=&(\partial_{\mu}\gamma_{\mu}-m)(\partial_{\mu}\gamma_{\mu}+m)G_D(z,z^{\prime})\nonumber\\&=&\gamma^1 \delta^{\prime}(x-x^{\prime})\delta(y-y^{\prime})+\gamma^2 \delta(x-x^{\prime})\delta^{\prime}(y-y^{\prime})-m\delta(x-x^{\prime})\delta(y-y^{\prime})\,.
\end{eqnarray}

\begin{figure} [tbp]
\centering
\leavevmode
\epsfxsize=8cm
\bigskip
\epsfbox{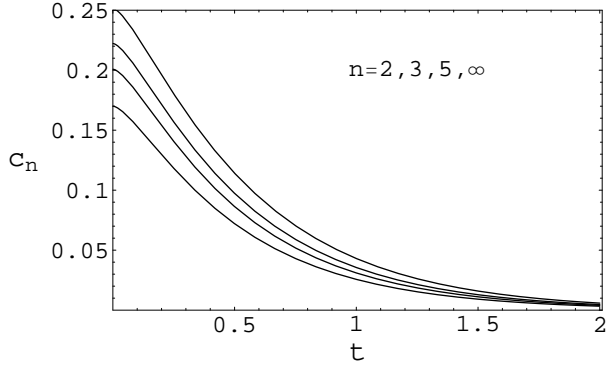}
\caption{The $c_n$ functions for a Dirac field and, from top to bottom,  $n=2$, $3$, $5$, and $n\rightarrow \infty$.}
\label{fyfy}
\end{figure}

 We define the function $T$ as the difference
\begin{equation}
G_D(z,z^{\prime})-\mathbb I ~m~G_S(z,z^{\prime})=T(z,z^{\prime})\,.
\end{equation}
The sum of the diagonal components and the difference of the off diagonal ones in $T$ satisfy the homogeneous Helmholtz equation according to (\ref{gsgd}). Thus, they are non zero only if they are not bounded at $L_1$ or $L_2$. According to (\ref{choa}) the type of divergence which is possible for the fermionic Green function is of the same type than the ones of $S_1(z)$ and $S_1(Rz)$ defined in section 3.1.1. Then we can expand 
\begin{equation}
T(z,z^{\prime})=\sigma_1 p(z,z^{\prime})+\sigma_3 q(z,z^{\prime})+\mathbb I w^{\dagger}(z)f(z^{\prime})+\sigma_2 w^{\dagger}(z)g(z^{\prime}) 
\end{equation}
where $p$, $q$, $w$, $f$, $g$ are unknown functions, and $w(z)=\begin{pmatrix}S_1(z)\\S_1(Rz)\end{pmatrix}$.
Having reached to this anzats, we can now impose the hermiticity condition $\gamma^3 G_D^\dagger \gamma^3=G_D$, where $\gamma^3=i \gamma^1 \gamma^2$, and the Dirac equation (\ref{deq}) for $G_D=\mathbb I ~m~G_S+T$. This gives enough equations to determine the unknown functions in terms of quantities related to the scalar Green function. Making use of the equations found for the scalar case, we arrive at the following result for the trace 
\begin{equation}
\textrm{tr}G_D=2 m\, \textrm{tr}G_S -16 \pi a (1-a)\frac{X_1}{\beta_1}\,.
\end{equation} 
Writing this in terms of $u$ and $u^\prime$ and using  (\ref{pero}) we have
\begin{equation}
w_a(t)=L \frac{d(\log Z)}{dL}=-\int_t^\infty dy \,\,y \left( \frac{(1-2a)\,u(y) +y\,u^{\prime}(y)}{y\,(1+u^2(y))}\right)^2\,,\label{pejot}
\end{equation} 
with $t=m L$. Comparing this expression with the corresponding one for the scalar (\ref{kuu}), it suggests that we could write it simply as
\begin{equation}
w_{a }(t)=-\int_{x}^{\infty }y\,v^{2}(y)\,dy \,,  \label{uni}
\end{equation} 
where
\begin{equation}
v(t)=\frac{(1-2a)\,u(t) +t\,u^{\prime}(t)}{t\,(1+u^2(t))}\,,\label{tiro}
\end{equation}
and search for a differential equation for $v$. Surprisingly, this variable satisfies a similar differential equation of Painlev\'e V type as the one satisfied by $u$,   
\begin{equation}
v^{\prime \prime }+\frac{1}{t}v^{\prime }=-\frac{
v}{1-v^{2}} v^{\prime 2}+v-v^{3}+\frac{4 a^{2}}{t^{2}}\frac{v}{1-v^{2}}\,.  \label{q}
\end{equation}

\begin{figure} [tbp]
\centering
\leavevmode
\epsfysize=5.5cm
\bigskip
\epsfbox{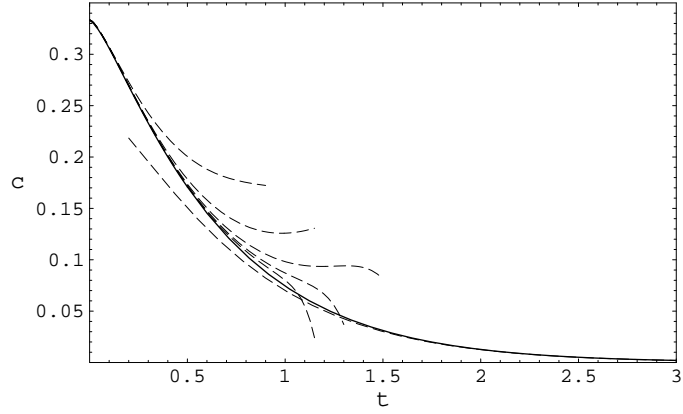} 
\caption{The $c$ function for a Dirac field. Also shown with dashed lines are the long distance leading term and the series expansions around the origin up to orders $t^2$, $t^4$, $t^6$, $t^8$, and $t^{10}$.}
\label{fity}
\end{figure}

This follows from (\ref{tiro}) and the equation (\ref{ecdif}) for $u$. The transformation (\ref{tiro}), which maps functions satisfying Painlev\'e equations to functions satisfying different Painlev\'e equations is one particular case of the transformation group studied in \cite{rims2}. The boundary conditions follows from the ones for $u$,
\begin{eqnarray}
v(t) &\rightarrow &\frac{2}{\pi} \sin ( \pi a) K_{2a} (t)\,\,\,\,\,\,\,\, \textrm{as} \,\,\,\, t\rightarrow \infty \,,\label{large}\\
v(t)&\rightarrow& -2a \,(\log t+\,\kappa_D)\,\,\,\,\,\,\,\textrm{as}\,\,\,\,t\rightarrow 0 \label{fermion0}\,,
\end{eqnarray}
where   $
\kappa_D=\rm -log(2)+2{\gamma}_E+\frac{\psi[a]+\psi[-a]}{2}
$.

Thus, (\ref{uni}), (\ref{q}) and (\ref{large}) give the exact value of $c_n(t)$ and $c(t)$,
\begin{eqnarray}
c_n(t)&=&\frac{1}{1-n}\sum_{k=-(n-1)/2}^{(n-1)/2}w_{k/n}(t)\,.
\label{doblew}
\\
c(t)&=&\int_0^{\infty}db \frac{\pi}{\sinh(\pi b)^2}w_{-i b}(t)\,.\label{uuy}
\end{eqnarray}
The eq. (\ref{uuy}) follows from (\ref{vein}) using $\lambda=e^{i 2\pi a}$ and $a=-i b$, with $b\in (0,\infty)$. 
The functions $c_n(t)$ are shown in figure \ref{fyfy} for some values of $n$.

 The leading long distance terms on these functions read
\begin{eqnarray}
c_n(t) &\sim& \frac{n}{n-1}\frac{e^{-2t}}{2\pi}+{\cal O}(\frac{e^{-2t}}{t})\,,\label{distan}\\
 c(t)&\sim& \frac{1}{2} t K_1(2 t)\,.\label{lardisbe}
\end{eqnarray} 
Eq. (\ref{lardisbe}) gives twice the corresponding term for a real scalar. Remarkably, it has been recently shown using the form factor approach, that the large distance leading term behavior $c(L)\simeq 1/4\sum_i m_i L K_1(2 m_i L)$, where the sum is over the different particles of the theory counted with their multiplicity, also holds for the integrable interacting theories \cite{ccd}, and also out of integrability \cite{dy}.   

In order to improve the short distance expansion we can expand $v_{a }(t)$ by
a direct use of the differential equations close to the conformal limit. We have the series
solution of (\ref{q}) around the origin
\begin{equation}
v_{a }(t)=\sum_{m=0}^{\infty }t^{2m}\sum_{n=0}^{2m+1}f_{m,n}\log
^{n}(t)\,.
\end{equation}
The full expansion requires only the knowledge of the constant term $
f_{0,0}=\kappa_D$. The first terms read
\begin{eqnarray}
v_{a }(t) &=&-2 a \log (t)+\kappa_D+t^{2}\left( \frac{1}{4}\left( 2 a
-8 a ^{3}+\kappa_D-8 a ^{2}\kappa_D-4 a \kappa_D^{2}-\kappa_D^{3}\right) +\right.   
\\
&&+\frac{1}{2}\left( - a +8 a ^{3}+8 a ^{2}\kappa_D+3 a
\kappa_D^{2}\right) \log (t) \left. +\left( -4 a ^{3}-3 a ^{2}\kappa_D\right) \log
^{2}(t)+2 a ^{3}\log ^{3}(t)\right) +\mathcal{O}(t^{4}) \,.\nonumber 
\end{eqnarray}
 The integration constant for $
w_{a }$ is given by the conformal limit $w_{a
}(0)=-2 a ^{2}$. The short distance expansion of the c-function follows from this expansion through (\ref{uni}) and (\ref{uuy}). It has the general form 
\begin{equation}
c(t)=  \sum_{m=0}^{\infty }t^{2m}\sum_{n=0}^{2m}c_{m,n}\log
^{n}(t)\,,
\end{equation}
where the $c_{m,n}$ are functions of the $f_{m,n}$.
The first terms are
\begin{equation}
c(t)\sim\frac{1}{3}- \frac{t^2\,{\log (t)}^2}{3}+ t^2\,\log (t)\left( \frac{1}{3} +\frac{-1 - 6\,\gamma_E + 6\,\log (2)}{9} \right) \, - 0.163494\, \,t^2 \,.\label{expi}
\end{equation}
  The expansion up to order $t^{10}$ is plotted in figure \ref{fity}. The series does not seem to converge for $t>1$. For $c_n$ the expansion starts with the leading terms
  \begin{equation}
 c_n(t)\sim \frac{n+1}{6n}-\frac{n+1}{6n}t^2 \log t^2+{\cal O}(t^2 \log(t))\,.
\end{equation}

\subsubsection{Summary for one interval: Dirac vs. scalar fields}

\begin{figure} [tb]
\centering
\leavevmode
\epsfysize=5cm
\bigskip
\epsfbox{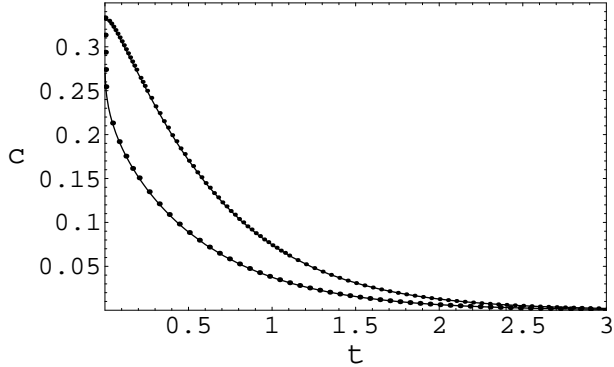}
\caption{Numerical vs. analytical results for the scalar (the lower curve) and Dirac (top curve) c-functions.}
\label{pietro}
\end{figure}

Interestingly enough, when it comes to the entanglement entropy, the Dirac and the scalar $c$ functions depend on one and the same differential equation. In order to see this, we only have to change variables. We summarize the results for a complex scalar and a Dirac field in the following way 
\begin{eqnarray}
c_D&=&\int_0^{\infty}db \frac{\pi}{\sinh(\pi b)^2}w_{D}\,,\\
c_S&=&\int_0^{\infty}db \frac{\pi}{\cosh(\pi b)^2}w_{S}\,,
\end{eqnarray}
where
\begin{eqnarray}
w_{(D,\,S)}&=&\int_{t}^{\infty}dy\,\,y\,\,u_{(D,\,S)}^{2}(y)  \,,\\
u_{(D,\,S)}{^{\prime \prime }}+\frac{1}{t}u_{(D,\,S)}^{\prime} &=&\frac{
u_{(D,\,S)}}{1+u_{(D,\,S)}^{2}}\left( u_{(D,\,S)}^{\prime}\right)
^{2}+u_{(D,\,S)}\left( 1+u_{(D,\,S)}^{2}\right) -\frac{4b^{2}}{t^{2}}\frac{u_{(D,\,S)}}{1+u_{(D,\,S)}^{2}}\,. 
\end{eqnarray}
The differences between the fermionic and scalar cases are exclusively due to the boundary conditions 
\begin{eqnarray}
u_D(t) &\rightarrow &\frac{2}{\pi} \sinh (b \pi) K_{i 2 b} (t)\,\,\,\,\,\,\,\, \textrm{as} \,\,\,\, t\rightarrow \infty \,\\
u_{D}(t)&\rightarrow& -2 b \,(\log t+\,\kappa_{D})\,\,\,\,\,\,\,\textrm{as}\,\,\,\,t\rightarrow 0 \\
u_{S}(t)&\rightarrow &\frac{2}{\pi} \cosh (b \pi) K_{i 2 b} (t)\,\,\,\,\,\,\,\, \textrm{as} \,\,\,\, t\rightarrow \infty \\
u_{S}(t) &\rightarrow & \frac{-1}{t \,(\log t+\kappa_S)}\;\;\;\;\;\textrm{as}\;\;t\to 0\,,
\end{eqnarray}
where $\kappa_D=-\log(2)+2\gamma_E+\frac{\psi[ib]+\psi[-ib]}{2}$ and $\kappa_S=-\log(2)+2\gamma_E+\frac{\psi[1/2+ib]+\psi[1/2-ib]}{2}$.

We show in figure \ref{pietro} a comparison of the exact analytical c-functions for a scalar and Dirac fields, along with the results from the lattice numerical simulations (see section 3.1.8). The bosonic $c$-function has the same central charge at the conformal point as the Dirac field, and thus they both tend to $1/3$ at the origin. The scalar function then rapidly goes to half the Dirac one (that is, approaches the c functions corresponding to a Majorana field). Note also the very different behavior at the origin. The bosonic c-function has a $1/\log (t)$ term that can be ascribed to the zero mode which is present at the conformal point.

\subsubsection{Entropy saturation at long distance}
 At large $t=L m$ the function $c(t)=L d S(L)/dL$ quickly goes to zero, and the entropy stops to grow. For this saturation limit one has
\begin{equation}
S(\infty)=\int_{\epsilon m}^\infty dy \,\,\frac{c(y)}{y} =-c(0) \log(m \epsilon)+\textrm{const}\,.
\end{equation}
This result about the dependence on $\log (m)$ of the saturation constant for the entropy was established with all generality in \cite{cc}. For the scalar and the Dirac field we have $c(0)=1/3$. For the scalar, a subleading term $\frac{1}{2}\log(-\log(m))$ term is also present.

\subsubsection{Dirac field: bosonization and the massless multicomponent case}

\begin{figure} [tbp]
\centering
\leavevmode
\epsfxsize=10cm
\bigskip
\epsfbox{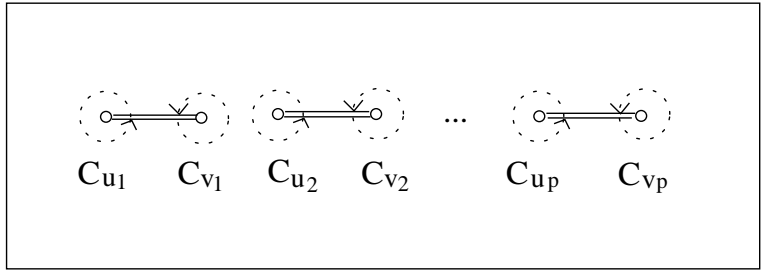}
\caption{The plane cut along the intervals $(u_i,v_i)$ with $i=1,..p$}
\label{fihi}
\end{figure}

Consider a massless Dirac field and a general set $V$ consisting in a collection of
disjoint intervals $ (u_{i},v_{i})$, $i=1,...,p$ (figure \ref{fihi}).
 In order to calculate $S(V)$ and $S_n(V)$ we follow section 2.1.2.  Accordingly, the problem is reduced to $n$ decoupled
fields $\Psi^{k}$ living on a single plane. These fields are multivalued,
since when encircling $C_{u_{i}}$ or $C_{v_{i}}$ they are multiplied
by $e^{i\frac{k}{n}2\pi }$ or $e^{-i\frac{k}{n}2\pi }$, respectively.

That multivaluedness can be disposed of, at the expense of
coupling singled-valued fields $\Psi^{k}$ to an external gauge
field which is a pure gauge everywhere, except at the points $u_{i}$ and
$v_{i}$ where it is vortex-like. Thus we arrive to the Lagrangian density
\begin{equation}
{\cal L}_{k}=\bar{\Psi}^{k}\gamma ^{\mu }\left( \partial _{\mu }+i\,A_{\mu
}^{k}\right) \Psi ^{k}+m\bar{\Psi}^{k}\Psi ^{k}\,.
\end{equation}
The reverse step would be to get rid of the gauge field $A_{\mu }$
by performing a singular gauge transformation 
\begin{equation}
\Psi^{k}(x)\to e^{-i\int_{x_{0}}^{x}dx^{^{\prime }\mu }A_{\mu }^{k}(x^{^{\prime }})}\Psi
^{k}\left( x\right) \;,
\end{equation}
(where $x_{0}$ is an arbitrary fixed point). Since the transformation
is singular, one goes back to a multivalued field. In order to reproduce the boundary conditions on $\Psi^{k}$,
we should have 
\begin{equation}
\oint_{C_{u_{i}}}dx^{\mu }A_{\mu }^{k}(x) =-\frac{2 \pi k}{n} \,,  
\hspace{1.2cm}
\oint_{C_{v_{i}}}dx^{\mu }A_{\mu }^{k}(x) =\frac{2 \pi k}{n} \,.
\label{90}
\end{equation}

Equations in (\ref{90}) hold for any two circuits
$C_{u_{i}}$ and $C_{v_{i}}$  containing $u_{i}$ and $v_{i}$
respectively. Thus
\begin{equation}
\epsilon ^{\mu \nu }\partial _{\nu}A_{\mu }^{k}(x)=2\pi \frac{k}{n}
\sum_{i=1}^{p}\big[ \delta (x-u_{i})-\delta (x-v_{i})\big] \,,  \label{aa}
\end{equation}
where the presence of a vortex-antivortex pair for each $k$ and each interval is
explicit.  
Then, the partition function  can be obtained as vacuum expectation values in the free Dirac
theory 
\begin{equation}
Z[e^{i 2 \pi k/n}]=\left\langle e^{i\int A_{\mu }^{k}j_{k}^{\mu }d^{2}x}\right\rangle \,,
\label{fj}
\end{equation}
where $j_{k}^{\mu }$ is the Dirac current, $A_{\mu }^{k}$ satisfies
(\ref {aa}), and we adopted a normalization such that $\left\langle 1
\right\rangle = 1$.

In order to evaluate (\ref{fj}), it is quite convenient to use the
bosonization technique in two dimensions \cite{fermion}. 
One expresses the fermionic current in terms of a dual scalar field $\phi$ as 
\begin{equation}
j_{k}^{\mu }\to \frac{1}{\sqrt{\pi }}\epsilon ^{\mu \nu }\partial_{\nu}\phi\,.
\end{equation}
Then, the functional becomes
\begin{equation}
Z[e^{i 2 \pi k/n}]=\left\langle e^{i\int A_{\mu }^{k}\frac{1}{\sqrt{\pi }}\epsilon ^{\mu
\nu }\partial _{\nu}\phi d^{2}x}\right\rangle =\left\langle e^{-i\sqrt{4\pi }
\frac{k}{n}\sum_{i=1}^{p}\left( \phi (u_{i})-\phi (v_{i})\right)
}\right\rangle \,,
\label{zz}
\end{equation}
where the vacuum expectation values correspond to the scalar
theory.  For the massless fermion this is simply a free massless scalar
\begin{equation}
{\cal L}_\phi=\frac{1}{2} \partial_\mu \phi \partial ^\mu \phi\,.
\end{equation}
 Since ${\cal L}_\phi$ is quadratic
\begin{equation}
\left\langle e^{-i\int f(x)\phi (x)d^{2}x}\right\rangle =e^{-\frac{1}{2}\int
f(x)G(x-y)f(y)d^{2}xd^{2}y}\,,
\end{equation}
with the correlator 
\begin{equation}
G(x-y)=-\frac{1}{2\pi }\log \left| x-y\right| \,.
\end{equation}
It follows that (\ref{zz}) can be written as
\begin{eqnarray}
\log Z[e^{i 2 \pi k/n}] &=&-\frac{2k^{2}}{n^{2}}\Xi \left( V\right) \,, \\
\Xi \left( V\right) &=&\sum_{i,j}\log
\left| u_{i}-v_{j}\right|-\sum_{i<j}\log \left|
u_{i}-u_{j}\right| -\sum_{i<j}\log \left| v_{i}-v_{j}\right|  -p\log \varepsilon  \,.
\end{eqnarray}
Here $\varepsilon $ is a cutoff introduced to split the coincidence points, $
\left| u_{i}-u_{i}\right| $, $\left| v_{i}-v_{i}\right| \to \varepsilon $. Summing
over $k$ and using (\ref{cheto}) and (\ref{alpha}) we obtain
\begin{eqnarray}
S_n &=& \frac{1}{6} \frac{n+1}{n}
 \Xi \left( V\right) \,,\label{popopo} \\
S &=&\frac{1}{3}\Xi \left( V\right) \,.  \label{et}
\end{eqnarray}
This formula first appeared in \cite{cc} as a proposal for the entanglement
entropy of disjoint intervals for any conformal theory in two dimensions. It was later discovered that this is not the case \cite{cg}. However, it is the correct formula for the free massless fermion model. It is interesting that (\ref{et}) does not seem to hold for the a model dual to the free fermion given in terms of a compactified scalar \cite{rems}. The reason for this is probably that while the theories can be mapped to each other, this mapping is non-local, and the definition of what is the algebra of operators in region differ between these models \cite{furu}.

Equation (\ref{et}) has an interesting corollary: recalling the
definition (\ref{mutual}) for the mutual information, it follows that, for
non-intersecting sets $A$, $B$ and $C$
\begin{equation}
I(A,B\cup C)=I(A,B)+I(A,C)\,.
\end{equation}
That is, in contrast to the entropy, the mutual information is extensive for the massless free fermion. This curious property
does not hold in the massive case or in more dimensions \cite{remarks}. 

\subsubsection{Scalar and Dirac partition functions as Sine-Gordon correlators}
We can also use the bosonization technique to deal with (\ref{fj}) for the
partition function for the case of a massive fermion.
This gives place to an alternative way of obtaining the partition functions already studied in sections 3.1.1 and 3.1.2.
As in the massless case, we still  
have eq. (\ref{zz}) for the partition function. Now, however, the bosonization of the massive fermion theory
leads to a sine-Gordon theory at the free fermion point \cite{sinegordon}, with Lagrangian 
\begin{equation}
{\cal L}_\phi=\frac{1}{2}\left( \partial_\mu \phi \partial ^\mu \phi+\Lambda \cos(\sqrt{4 \pi}\phi)\right)\,,\label{sg}
\end{equation}
where $\Lambda$ is a dimensionful parameter. 
 
This allows us to identify the partition functions with sine-Gordon correlators of vertex operators. In the case of a single interval we have  
\begin{equation}
Z[e^{i 2 \pi a}] \simeq \left< V_{a}V_{-a} \right>\,,\label{fghj}
\end{equation}
where $V_{a}$ is $:\exp[i\sqrt{4\pi}a \phi]:$ and  $a\in (-1/2,1/2)$. These type of correlators have been studied in the literature, starting from the one for $a=1/2$ which is related to the Ising model spin and magnetization correlators \cite{ising}. The correlator (\ref{fghj}) has a long distance expansion in terms of intermediate particle states, called form factor expansion. This reads
\begin{eqnarray}
\left\langle :e^{i\sqrt{4\pi } a \phi (r)}::e^{i\sqrt{4\pi } a
^{\prime }\phi (0)}:\right\rangle &=&\sum_{n=0}^{\infty }\frac{1}{\left(
n!\right) ^{2}}\int_{0}^{\infty }du_{1}...du_{2n}\left( \prod_{i=1}^{2n}e^{-
\frac{mr}{2}\left( u_{i}+\frac{1}{u_{i}}\right) }\right) \nonumber \\
&&\times f_{a }(u_{1},...,u_{2n})f_{a ^{\prime
}}(u_{2n},...,u_{1})\,,  \label{hache1}
\end{eqnarray}
where $f_{a }(u_{1},...,u_{2n})$ is the form factor
\begin{equation}
f_{a }(u_{1},...,u_{2n})=(-1)^{n(n-1)/2}\left( \frac{\sin \left( \pi
a \right) }{i\pi }\right) ^{n}\left( \prod_{i=1}^{n}\left( \frac{u_{i+n}
}{u_{i}}\right) ^{a }\right) \times \Delta (u_{1},...,u_{2n})\,,
\end{equation}
and
\begin{equation}
\Delta (u_{1},...,u_{2n})=\frac{\prod_{i<j\leq n}\left( u_{i}-u_{j}\right)
\prod_{n+1\leq i<j}\left( u_{i}-u_{j}\right) }{\prod_{r=1}^{n}
\prod_{s=n+1}^{2n}\left( u_{r}+u_{s}\right) }\,.
\end{equation}
The first term on this expansion gives place to the long distance behavior (\ref{lardisbe}) for the entropy \cite{fermion,ccd}. It can be shown that this series may be summed as a Fredhlom determinant and satisfies the Painlev\'e eqs. (\ref{pejot}-\ref{fermion0}) \cite{suma}.  This proof follows a very different route than the one in sections 3.1.1 and 3.1.2. 
The form factor expansion and integrable field theory techniques have been successfully used to approximate the entanglement entropy in one dimensional integrable models \cite{ccd,d1,dc}.

For a scalar field we also have an expression of the partition function in terms of sine-Gordon correlators,
\begin{equation}
Z[e^{i 2 \pi a}] \simeq \frac{1}{\left< V_{a}V_{1-a} \right>}\,,
\end{equation}
with $a\in(0,1)$. This can be checked directly using the Painlev\'e equations, and (\ref{fghj}). It also follows from the form factor expansion for the scalar case \cite{mussardo}.

\subsubsection{Dirac field: small mass expansion in the real time approach}

In order to compute the entropy for the multicomponent massive case, we find convenient to use the real time approach. 
First, we express the real time approach entropy formula (\ref{tiuno}) in terms of the resolvent $R=(C-1/2+\beta)^{-1}$ as
\begin{equation}
S(V)=-\int^\infty_{1/2} d\beta\, \textrm{tr}\left[\left(\beta-1/2\right) \left(R(\beta)-R(-\beta)\right)-\frac{2\beta}{\beta+1/2}\right]\,.\label{fas}
\end{equation}  
The correlator for a two dimensional Dirac field reads 
\begin{equation}
C(x,y)=\frac{1}{2} \delta(x-y) +\frac{m}{2\pi} K_0(m|x-y|) \gamma^0  +\frac{i m}{2 \pi} \, K_1(m (x-y)) \gamma^3  \,,\label{crosta}
\end{equation}
Where $\gamma^0$ and $\gamma^1$ are the Dirac matrices, and $\gamma^3=\gamma^0 \gamma^1$. 
In the massless limit this gives 
\begin{equation}
C_0 (x,y)=\frac{1}{2} \delta(x-y)\, 
 1+ \frac{i}{2\pi}\frac{1}{x-y} \gamma^3\,. \label{hora}\\
\end{equation} 
Fortunately, the resolvent for this integral operator inside a region formed by $n$ disjoint intervals $(u_i,v_i)$ is known from the theory of singular integral equations  \cite{reso}. We have  
\begin{equation}       
R^0(\beta) =\left(\beta^2-1/4 \right)^{-1}
\left(\beta\,\delta(x-y)\, 
+\frac{i \gamma^3}{2\pi}   \frac{e^{-\frac{i}{2\pi}\gamma^3 \log\left(\frac{\beta-1/2}{\beta+1/2}\right)\, (z(x)-z(y)) }}{x-y}
\right)\,,\label{reso}
\end{equation}
where
\begin{equation}
z(x)=\log\left(-\frac{\prod_{i=1}^n (x-u_i)}{\prod_{i=1}^n (x-v_i)}\right) \,. \label{bfbf}
\end{equation}
With this resolvent at hand we are in position to compute the entropy for the massless case and to make expansions for the massive case \cite{futuro}.

 In (\ref{fas}) the term proportional to the identity cancels with the corresponding one in the resolvent (\ref{reso}). Then we have, 
\begin{equation}
S_{m=0}(V)=-\frac{2}{\pi}\int^\infty_{1/2} d\beta\, \int_V dx\, \lim_{y\rightarrow x}  \frac{\sin\left[ \frac{1}{2\pi} \log\left(\frac{\beta-1/2}{\beta+1/2}\right)\, (z(x)-z(y)) \right]}{(\beta+1/2)\,(x-y)}\,.
\end{equation}  
 Integrating over $\beta$ first, this gives
\begin{eqnarray}
S_{m=0}(V)&=&2\int_V dx\, \lim_{y\rightarrow x}  \frac{\frac{z(x)-z(y)}{2}\coth((z(x)-z(y))/2)-1}{(x-y)(z(x)-z(y))}=\frac{1}{6}\int_V dx\,\sum_{i=1}^n \left(\frac{1}{x-u_i}-\frac{1}{x-v_i}\right)\nonumber\\
&\,&\hspace{1cm}=\frac{1}{3} \left( \sum_{i,j}\log|v_i-u_i|-\sum_{i<j} \log|u_i-u_j| -\sum_{i<j} \log|v_i-v_j|-n \log \epsilon \right)\,.\label{sesen}
\end{eqnarray}
 This corresponds to the eq. (\ref{popopo}) obtained with the Euclidean method \cite{fermion}.

The short distance expansion of the correlator up to second order in the field mass is $C=C_0+C_1+C_2+...$ where
\begin{eqnarray}
C_1 (x,y)&=&-\frac{m}{2\pi}\left(\gamma_E+\log\left(\frac{m|x-y|}{2}\right)\right) \gamma^0 \,,\\
C_2 (x,y)&=&\frac{i m^2}{4\pi} (x-y) \left(\gamma_E -\frac{1}{2} +\log\left(\frac{m|x-y|}{2}\right) \right) \gamma^3\,.
\end{eqnarray}
The perturbative expansion of the resolvent for small mass is 
\begin{equation}
R_V(\beta)=R_V^0(\beta)-R_V^0(\beta)C_1 R_V^0(\beta)-R_V^0(\beta)C_2 R_V^0(\beta)+R_V^0(\beta)C_1 R_V^0(\beta)C_1 R_V^0(\beta)-...\,.
\label{financiar}
\end{equation}
Using this in (\ref{fas}) we get (see \cite{futuro} for details)
\begin{eqnarray}
S(V)&=&S_{m=0}(V)-\frac{m^2 L_t^2}{6}  \log^2(m)+\log(m)\left(\frac{2\log(2)-2\gamma_E-1}{6} m^2 L_t^2\right. \\
 &&\hspace{1cm}\left. - 2 m^2  \int_V dx\,\int_V dy\, \log|x-y|\, \delta(z(x)+z(y))\right)+{\cal O}(m^2 \log(m)^0) \nonumber \,,
\end{eqnarray}
where $L_t=\sum_i (v_i-u_i)$ is the total length of the intervals. For a single interval this coincides with the corresponding terms in the expansion (\ref{expi}) obtained with the help of the differential equation. The last term is the first one which gives a non-extensive mutual information \cite{remarks}.

\subsubsection{Lattice Hamiltonian and correlators in one  dimension}

In this section we describe the Hamiltonian and correlators which can be used in a one dimensional lattice 
to calculate numerically the entanglement entropy. The numerical results of the $c$ functions for a scalar and a Dirac field are compared with the analytical ones in figure \ref{pietro}, showing a perfect accord. 

\noindent\textbf{Scalar}

\noindent We take the lattice Hamiltonian for a real massive scalar as 
\begin{equation}
{\cal H}=\frac{1}{2}\sum_{n=-\infty}^{\infty}\left( \pi _{n}^{2}+(\phi _{n+1}-\phi
_{n})^{2}+\,m^2\phi _{n}^{2}\right) \,.
\end{equation}
 We have 
set the lattice spacing to one. 
The correlators (\ref{x}) and (\ref{p}) are  
\begin{eqnarray}
\langle \phi_{n}\phi_{m}\rangle &=& \int_{-\pi}^{\pi}dx\frac{e^{ ix(m-n)}}{4\pi\sqrt{
m^2+2(1-\cos ( x))}}\,, \\
\langle \pi_{n}\pi_{m}\rangle &=& \int_{-\pi}^{\pi}dx \frac{1}{4\pi}e^{ i x(m-n)}\sqrt{m^2+2(1-\cos ( x))}\,.
\end{eqnarray}

\noindent\textbf{Dirac fermions}

\noindent We take the Hamiltonian for a Dirac fermion
discretized on the lattice 
\begin{equation}
{\cal H}=\sum_{n=-\infty}^\infty\left( -\frac{i}{2}\left(\Psi^\dagger_{n} \gamma^0\gamma^1(\Psi_{n+1}-\Psi_{n})-\textrm{h.c.}\right)+m  \Psi_{n}^\dagger\gamma^0 \Psi_{n}\right) \,. 
\end{equation}
The two dimensional matrices $\gamma^0$ and $\gamma^1$ can be taken at will while satisfying the fundamental relations for the Dirac matrices (i.e. $\gamma^0=\sigma_1$ and $\gamma^1=i \sigma_2$).  
The correlator (\ref{ffff}) is
\begin{equation}
\langle \Psi^\dagger_{i}\Psi_{j}\rangle=\frac{1}{2}\delta_{i,j}+\int_{-\pi}^\pi dx\,\frac{m \gamma^0+\sin( x)\gamma^0 \gamma^1}{4 \pi \sqrt{m^2+\sin^2( x)}} e^{ i x (i-j)}\,.
\end{equation}
Due to the fermion doubling on the lattice, one has to divide the lattice results by $2$ in two dimensions in order to get the entropy corresponding to a Dirac field in the continuum limit.

\subsection{Two dimensions: logarithmic term for polygonal sets}
\begin{figure} [tbp]
\centering
\leavevmode
\epsfysize=5.3cm
\bigskip
\epsfbox{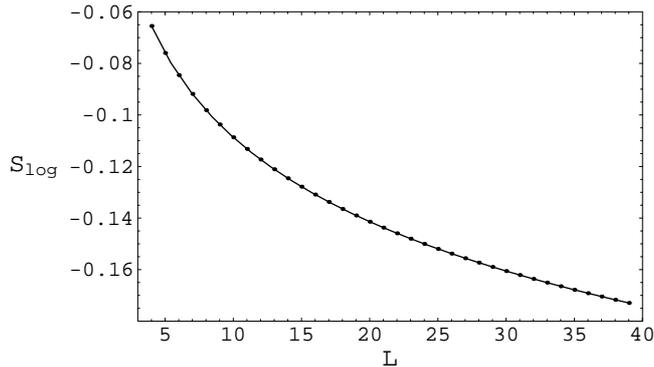}
\caption{The logarithmic term for the entropy of a massless scalar, where $V$ is a square of side $L$. Here we have plotted the results of the numerical evaluations of the entanglement entropy on the lattice, where we have subtracted the linear and constant terms in a fit of the form $S=c_0+c_1 L-s \log L$. The solid curve is $-0.0472\,\log L$, showing perfect accord with the numerical data.}
\label{haha}
\end{figure}

In the continuum limit described by a quantum field theory the entanglement entropy is divergent due to the presence of an unbounded number of local degrees of freedom.  The singularities structure is encoded in the expansion (\ref{div}) where the dimensionless coefficient $g_0(V)$ of the logarithmic term is particularly relevant since it is the only one universal. 
Logarithmic divergent terms in the entropy have been previously found in four dimensional black hole space-times \cite{dos}. They are present generically in even spacetime dimensions ($d$ odd) for sets with smooth curved boundaries. This follows from the heat kernel expansion for conical manifolds with smooth singularity surface (see section 3.3.3). 
In \cite{log} and \cite{log1} it was shown that there is also a logarithmic term in $d=2$ for sets $V$ with non-smooth boundary (see also \cite{frolovangulo,frad}). The figure \ref{haha} shows the logarithmic term for a square in a two dimensional lattice. 
Since $g_0(V)$ is dimensionless, extensive and local on the boundary, for $V$ a spatial polygonal set, it must be of the form
\begin{equation}
g_0(V)=\sum_{v_i} s(x_i)\,,
\end{equation}       
where the sum is over all vertices $v_i$ and  $x_i$ is the vertex angle. On general grounds one also expects point-like vertex induced logarithmic  terms in any dimensions.

The alpha-entropies also contain a logarithmic term 
\begin{equation}
\left.S_n\right|_{\log}=\sum_{v_i} s_n(x_i) \log(\epsilon \Lambda)\,, \label{cinco}
\end{equation} 
analogous to the one in the entropy, with $
s(x)=\lim_{n\rightarrow 1} s_n(x)$.
In eq. (\ref{cinco}) $\Lambda$ is a parameter with the dimensions of an energy, depending on $V$ and on the particular  theory. For a massless field it is the inverse of any typical dimension $R$ of $V$ and when the mass dominates, $MR\gg 1$, it can be taken  as $\Lambda=M$. 

In this section we review the analytic results on $s(x)$ obtained in \cite{log} and \cite{log1} for a free scalar and Dirac fields respectively, and show the results match with the numerical simulations on a two dimensional lattice.

\subsubsection{Scalar field}

In order to calculate the logarithmic coefficient $s_n$ for a free scalar we consider a plane angular sector as the simplest set with vertex contributions. We start studying the Green function as in 3.1.1.  The eigenfunctions of the Laplacian admit separation of variables, and by direct calculation of the radial component of the eigenfunctions, we are left with a two dimensional reduced problem. This later consists in finding the trace of the Green function on a sphere with a cut with particular boundary conditions on it inherited from the original problem. The Myers method, used to find the entropy of a scalar field in a plane with a cut in 3.1.1 is well suited to the present case as well. The study of the divergences structure of the Green function in the vicinity of the set boundary and the identification of the symmetries are again the basic ingredients. 
\smallskip

\noindent
\textbf{Dimensional reduction}

\begin{figure} [tbp]
\centering
\leavevmode
\epsfxsize=6cm
\bigskip
\epsfbox{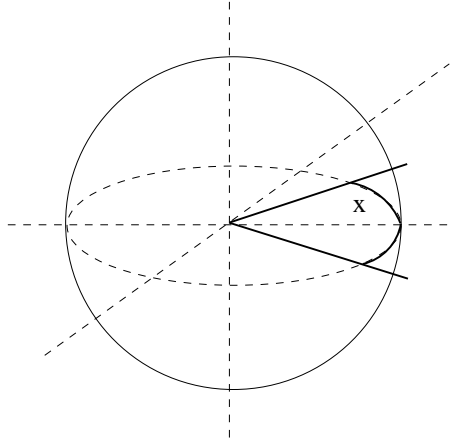}
\caption{Sphere with a cut at the intersection with the plane angular sector of angle $x$.}
\label{esfe}
\end{figure}

\noindent
We start considering the Green function $G(\vec{r}_1,\vec{r}_2)$ for a complex scalar of mass $M$ in three Euclidean dimensions subject to the boundary conditions (\ref{bc}). To be explicit, we have (see figure \ref{esfe})
\begin{eqnarray}
(-\Delta_{\vec{r}_1}+M^2) \,G(\vec{r}_1,\vec{r}_2)&=&\delta(\vec{r}_1-\vec{r}_2)\,,\\
\lim_{\varepsilon\rightarrow 0^+} G(\vec{r}_1+\varepsilon \hat{\eta},\vec{r}_2)&=&e^{i 2 \pi a}  \,\lim_{\varepsilon\rightarrow 0^+} G(\vec{r}_1-\varepsilon \hat{\eta},\vec{r}_2)\,, \hspace{1cm} r_1\in V\, ,\label{catorr}
\end{eqnarray} 
 where  $\hat{\eta}$ is orthogonal to the plane of $V$. This is related to the functional $Z$ according to 
\begin{equation}
\partial_{M^2}\textrm{log}Z[e^{i 2\pi a}]=-\int dr^3 G(\vec{r},\vec{r})\,.
\label{green}
\end{equation}

The Laplacian and the boundary conditions allow the separation of angular and radial equations in polar coordinates. Using standard methods we arrive at the expression 
\begin{equation}
G(\vec{r}_1,\vec{r}_2)=\sum_\nu \int d\lambda \frac{\lambda}{\lambda^2+M^2}\psi_{\nu}( \theta_1 , \varphi_1)\psi^*_{\nu}(\theta_2 , \varphi_2)\frac{J_{\frac{1}{2}+\nu}(\lambda r_1)J_{\frac{1}{2}+\nu}(\lambda r_2)}{\sqrt{r_1 r_2}}\,,
\label{ga}
\end{equation}
where $J$ is the Bessel function. Here the sum is over the normalized eigenvectors $\psi_{\nu}(\theta , \varphi)$ of the angular equation 
\begin{equation}
\Delta_{\Omega}\psi_{\nu}=-\nu(\nu+1)\psi_{\nu}\,,
\end{equation}
where $\Delta_{\Omega}$ is the Laplacian on the sphere, 
 $
\Delta_{\Omega}=\frac{1}{\sin \theta}\frac{\partial}{\partial \theta}(\sin\theta\frac{\partial}{\partial \theta})+\frac{1}{\sin^2\theta}\frac{\partial^2}{\partial \varphi^2}
$, 
with domain given by the functions satisfying the boundary conditions inherited from (\ref{catorr}). 
 The precise expressions for $\psi_{\nu}$ and $\nu$ are not relevant in what follows.

Taking the trace $\int dr^3 G(\vec{r},\vec{r})$ in eq. (\ref{ga}) gives (disregarding an unimportant divergent constant)
\begin{equation}
\partial_{M^2}\log Z=-\frac{1}{2M^2}\sum_{\nu}(\nu+1/2)=\frac{1}{2M^2}\textrm{tr}\sqrt{-\Delta_{\Omega}+\frac{1}{4}}\,.
\label{trsqrt}
\end{equation}
Though this expression is divergent, the piece we are interested in, which is the one dependent on the angle $x$, is finite. 
 To proceed, it is convenient to express the trace of the square root of the operator in (\ref{trsqrt}) in terms of the corresponding resolvent.  We have the identity \cite{seeley}  
\begin{equation}
\textrm{tr}  \sqrt{-\Delta_{\Omega}+\frac{1}{4}}=\frac{1}{\pi}\,\int_0^{\infty}\lambda^{\frac{1}{2}}\,\,\textrm{tr}\frac{1}{\Delta_{\Omega}-\frac{1}{4}-\lambda}\, d\lambda 
\,.
\label{tr}
\end{equation}

\smallskip 

\noindent\textbf{Green function on a sphere with a cut}

\noindent The problem is now reduced to the calculation of the trace of the two dimensional Green function on a sphere with a cut of angle $x$, where the boundary conditions inherited from (\ref{catorr}) are imposed. 
 
 This problem is the analogous on the sphere to the one solved for the plane in 3.1.2.
 Following the Myers method step by step as in 3.1.2,  we find the analytic expression for the trace of the Green function as a solution of a system of ordinary differential equations. The details of the derivation are given in \cite{log}. Explicitly we find   
\begin{equation}
\textrm{tr}G_s^{(2)}=8\pi (1-a)a\int_{x}^{\pi}H(y)dy\,.\label{quince}
\end{equation}

 \begin{figure} [tb]
\centering
\leavevmode
\epsfysize=5.3cm
\bigskip
\epsfbox{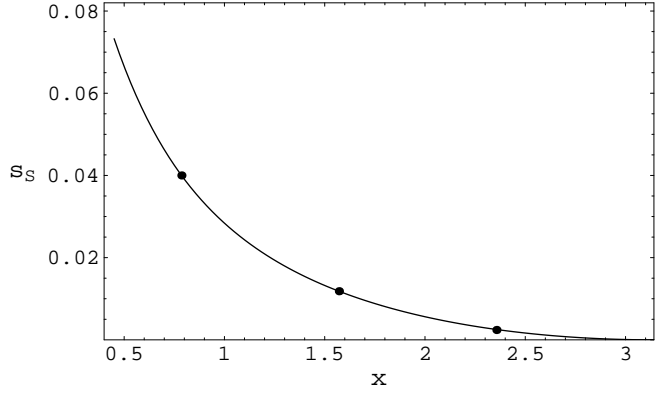}
\caption{Coefficient of the vertex induced logarithmic term in the entropy for a scalar field as a function of the vertex angle $x$. The three points shown correspond to numerical evaluations on the lattice for $x=\pi/4$, $\pi/2$, and $3 \pi/4$.}
\label{ryry}
\end{figure}
 
Here the function $H(x)$ is the solution of the following set of ordinary non linear differential equations (we omit the subscript $a$ and the dependence on $x$ of the variables for notational convenience)
\begin{eqnarray} 
H' &=& - \frac{m}{2}\,\left( b\,B_2 +c\, B_1 + 2\,u\,B_{12} \right)\label{hprima}\,,\\
X_1' &=& - m\,\left( b\,B_{12}+ u\,B_1 \right)\label{x1prima} \,,\\
X_2' &=& - m\,\left( c\,B_{12} + u\,B_2 \right) \label{x2prima}\,,\\
c' &=&- 2\,m\,\beta_2 \,u\, \csc(x)\,\sin(x/2) - c\,(1 - a)\,\csc(x)\,(1 + \cos(x))\label{cprima}\,, \\
b' &=& -2\,m\,\beta_1\,u\,\csc(x)\,\sin(x/2) - b\,a\,\csc(x)\,(1 + \cos(x)) \label{bprima} \,,\\
u'&=&-\frac{m}{2}\,\sec(x/2)\,(b\,\beta_2+ c\,\beta_1) + \frac{1}{2}\,u\,\tan \left(x/2 \right)\label{uprima} \,,
\end{eqnarray}
where $B_1$, $B_2$, $B_{12}$, $\beta_1$, $\beta_2$ are functions of $x$ given in terms of $H$, $X_1$, $X_2$, $c$, $b$, and $u$ by the following set of algebraic equations
\begin{eqnarray}
\frac{\cos(x/2)}{8 
\pi a(1 - a)}&=&\sin(x/2)\,H - m \left(\beta_1\, X_2 + \beta_2\, X_1\right) + 2\,m\,\cos(x/2)\, u\, B_{12}\,,\label{29} \\
\frac{\sin(x/2)}{8 \pi a(1 - a)}&=&-\cos(x/2)\,H - m\, \tan(x/2)\left(\beta_1 X_2 + \beta_2 X_1 \right) + 
  m\,\sin(x/2)(b B_2 + c B_1) \,,\label{30}\\
0&=& -m \sin(x/2)(c X_1
 -b X_2) + m \tan(x/2)(\beta_2 B_1- \beta_1 B_2)+(1-2a)\cos(x/2) B_{12}\,,\label{31}\\
0&=&-4a(a - 1) - m^2(4 - 8\beta_1\beta_2 + b c + 3u^2) \nonumber\\&&\hspace{2cm}- 
      4\cos(x)\left(a(a - 1) + m^2(u^2 + 1)\right) 
      + m^2\cos(2x)(b\,c - u^2) \label{4a}\,,\\
0&=&(2a - 1)u\cos(x/2)+m \tan(x/2)(\beta_1 c - b \beta_2)\,.\label{minuscula}
\end{eqnarray}
The boundary conditions at $x\rightarrow\pi$ are
\begin{eqnarray}
H(\pi)&=&0\,,\\
X_1(\pi)&=&\frac{ \Gamma(-a) \left( \cosh \left( \frac{\pi \mu}{2} \right) \textrm{Im} \left[ \psi \left( \frac{1}{2} + a + \frac{i\mu}{2} \right) \right] - \frac{\pi}{2} \sinh \left( \frac{\pi\mu}{2} \right)   \right)}{2^{2a}\mu  \left( \cos \left( 2 a \pi\right)+\cosh (\pi \mu)\right) \Gamma (1+a) \left| \Gamma \left( \frac{1}{2}-a+\frac{i\mu}{2}\right) \right|^2 }\,,\label{xx1}\\
X_2(\pi)&=&X_1(\pi)\left|_{\,a\rightarrow (1-a)} \right. \,,\label{xx2} \\
u(\pi)&=&0\,, \label{upi}\\
b	(\pi)&=&\frac{ 2^{1-2a} a (1-a) \left| \Gamma\left( \frac{1}{2} +a+\frac{i\mu}{2}\right)\right|^2}{m\Gamma^2 (1+a)}\,,\label{treintaytres}\\
c(\pi)&=&b(\pi)\left|_{\,a\rightarrow (1-a)}\right.  \,,\label{treintaydos}
\end{eqnarray}
where $\mu=\sqrt{4\,m^2-1}$ and $\psi$ is the digamma function. The meaning of the extra variables $B_1$, $B_2$, $B_{12}$, $X_1$, $X_2$, $u$, $b$, $c$, $\beta_1$ and $\beta_2$ is the same as in 3.1.2. The trace in (\ref{quince}) is regularized such that it vanishes when $x=\pi$, where there is no vertex point and no logarithmic term is present in the entropies. 

Gathering all the results together, using eqs. (\ref{cinco}) with $\Lambda =M$, (\ref{r1}), (\ref{trsqrt}), (\ref{tr}), and (\ref{quince}), we arrive at the final result for the logarithmic coefficients 
\begin{equation}
s_n(x)= \sum_{k=1}^{n-1}\frac{8\,k\,(n-k)}{n^2\,(n-1)}\int_{1/2}^{\infty}dm\,m\,(m^2-1/4)^{\frac{1}{2}}\int_{x}^{\pi}dy \,H_\frac{k}{n}(y,m)\,.
\label{finalescalar}
\end{equation}
and 
\begin{equation}
s(x)=\int_0^\infty dt \, \frac{16\pi (t^2+1/4)}{\cosh^2(\pi t)} \int_{1/2}^{\infty}dm\,m\,\left(m^2-1/4\right)^{\frac{1}{2}}\,\int_{x}^{\pi}dy \,H_{-it+1/2}(y,m) \,,
\label{finalescalar1}
\end{equation}
where the function $H$ is solution of the above set of ordinary non linear differential equations.

In  \cite{pain} the partition function for a Dirac fermion on the Poincar\'e disk with multiplicative boundary conditions imposed on a geodesic segment  has been written in terms of a solution of the Painlev\'e VI differential equation. Though we were not able to find an explicit relation, it is possible that our results for $H(x)$ could have also an expression in terms of solutions of these type of equations. 
The function $H$ also gives the exact entropy functions for a massive scalar and a spatial segment in $1+1$ dimensional de Sitter space \cite{log}.

An economic way to integrate the equations is to expand the functions involved in (\ref{hprima}-\ref{treintaydos}) in Taylor series around $x=\pi$, and obtain analytically the coefficients using the differential equations. Then the above integrals over $t$ and the mass can be done for each coefficient separately. With this method we have produced the curve of figure \ref{ryry}, which show $s_S(x)$ up to order $(x-\pi)^{14}$. Some of the coefficients are tabulated in table \ref{tatata}. In the picture are also plotted the values of $s_S$ for $x=\pi / 4$, $\pi /2$ and $3/4 \,\pi$ obtained by numerical simulations in the lattice. They show a perfect accord (around one percent error) with the analytical results. These particular values of the angle are the ones for which the coefficient can be calculated with very small error on a square lattice of limited size (in the present case it was $200\times 200$ points). The numerical methods consist of evaluating the entropy for a massless Dirac field  (see \cite{log}) for a given shape (square, triangle, etc.) and different overall size $\lambda$, and then fitting the result as $S=C_0+C_1\, \lambda+C_{-1}\, \lambda^{-1} +C_{-2} \,\lambda^{-2}- s_S \log(\lambda)$. 
 
The small angle limit of $s(x)$ relates the problem on the cut sphere with the corresponding one in the plane, treated in 3.1.1. One has from this mapping 
\begin{equation}
s(x)\sim\frac{\int_0^\infty dt\,\,c(t)}{\pi x}\,,\label{angulochico}
\end{equation}
 where $c$ is the entropic c-function for a scalar. This result can be understood in more general terms. This is explained in section 3.3.1.
\subsubsection{Dirac field}
The problem for a Dirac field and $V$ a plane angular sector can also be dimensionally reduced by separation of variables. Then, as in the scalar case, we have to calculate the trace of the Green function on a two dimensional sphere with a cut. This problem is then related to the scalar one already treated in 3.2.1. 

 The partition function for a Dirac field  $\Psi$ in three  dimensions is  
\begin{equation}
Z[e^{i 2 \pi a}]=\int {\cal D}\Psi^\dagger {\cal D}\Psi e^{-\int dr^3 \,\Psi^\dagger  {\cal D}_3 \Psi}\,,\label{eee}
\end{equation}
where ${\cal D}_3$ is the Dirac operator given by
\begin{equation}
{\cal D}_3=(\gamma^{i}\partial_i+\mu)\,,
\end{equation}
 and $\gamma^{i}=\sigma^{i}$ are the Pauli matrices and $\mu$ the mass. The boundary condition for the spinors is 
 \begin{equation}
\Psi^+(\vec{r})=e^{i 2\pi a} \Psi^-(\vec{r})\,,\hspace{2cm} \vec{r}\in V\,.\label{bounj}
\end{equation}
Here $\Psi^+$ and $\Psi^-$ are the limit values of the field on each of the sides the two dimensional angular sector $V$ has in three dimensions. 

The functional $Z$ is calculated exploiting the relation between the free energy and the Green function, 
\begin{equation}
\frac{d\log Z}{d\mu}=\textrm{Tr} \, G_D^{(3)}\,,\hspace{1cm}{\cal D}_3 G_D^{(3)}(\vec{r},\vec{r}^\prime) =\delta^3(\vec{r}-\vec{r}^\prime)\,.\label{chiri}
\end{equation}

\smallskip 

\noindent\textbf{Dimensional reduction}

\noindent The Dirac operator and the boundary conditions allow the separation of the angular and radial equations in polar coordinates. We use this fact in order to reduce the problem to one in two dimensions. 
In this coordinates ${\cal D}_3$ writes
\begin{equation}
	{\cal D}_3=\frac{D}{r}+\tilde{\gamma}^r\partial_r+\mu\,,
\hspace{1cm} D=(\tilde{\gamma}^{\theta}\partial_{\theta}+\tilde{\gamma}^{\phi}\partial_{\phi})\,,
\label{253}\end{equation}
and where the redefined gamma matrices $\tilde{\gamma}$ are 
\begin{equation}
\tilde{\gamma}^{\theta}=r\, \frac{\partial\theta}{\partial x_i} \sigma^i \,,\hspace{1cm}
\tilde{\gamma}^{\phi}=r\, \frac{\partial\phi}{\partial x_i} \sigma^i.\,,\hspace{1cm}
\tilde{\gamma}^{r}=\frac{\partial r}{\partial x_i} \sigma^i=-i\sin\theta\tilde{\gamma}^{\theta}\tilde{\gamma}^{\phi}\,.
\end{equation}
The dimensional reduction proceeds as in the scalar case, by integrating out the radial variables in the Green function. For details see \cite{log1}. We arrive at
\begin{equation}
\textrm{tr} \, G^{(3)}_D=-\frac{1}{2 \mu} \textrm{tr}\left|\tilde{\gamma}^rD-1/2\right|\,.\label{qwqw}
\end{equation}

Using (\ref{chiri}) and (\ref{253}) the logarithmic divergent contribution to the partition function  is
\begin{equation}
\log Z|_{\log}=-\frac{1}{2} \textrm{tr}\left|\tilde{\gamma}^rD-1/2\right| \log (\epsilon \mu)\,.\label{questa}
\end{equation}

The trace in (\ref{questa}) can be calculated using the integral representation in terms of the resolvent \cite{seeley} 
\begin{equation}
\textrm{tr}\left|\tilde{\gamma}^r D-1/2\right|=-\frac{1}{\pi}\int^{\infty}_{-\infty}dm\,m\, \textrm{tr}(i(\tilde{\gamma}^rD-1/2)+m)^{-1}\,.\label{refe1}
\end{equation}
The operator 
\begin{equation}
{\cal D}_2=i(\tilde{\gamma}^r D-1/2)+ m\label{dira}
\end{equation}
 is a two dimensional Dirac operator on the sphere where the parameter $m$ plays the role of a mass.

Thus, we have to find the trace of the Green function of a Dirac field on a two-dimensional sphere with a cut. This  
satisfies
\begin{equation}
{\cal D}_2 G_D^{(2)} =\sqrt{g}\, \,\delta^2(z-z^\prime)\,,
\end{equation}
and the boundary conditions (\ref{bounj}).

 \begin{figure} [tb]
\centering
\leavevmode
\epsfysize=5.3cm
\bigskip
\epsfbox{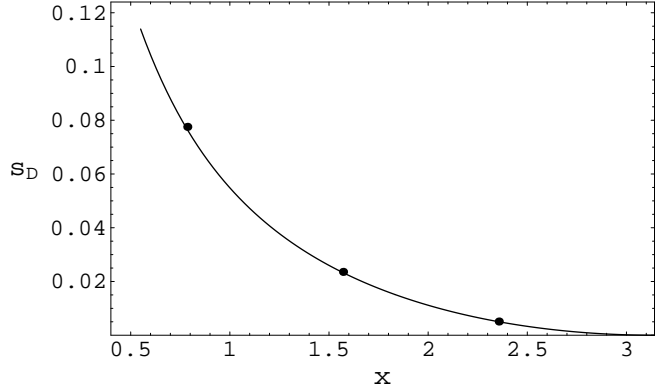}
\caption{Coefficient of the vertex induced logarithmic term in the entropy for a Dirac field as a function of the vertex angle $x$. The three points shown correspond to numerical evaluations on the lattice for $x=\pi/4$, $\pi/2$, and $3 \pi/4$.}
\label{ryo}
\end{figure}

\smallskip

\noindent\textbf{Relation between the scalar and Dirac Green functions}

\noindent As in the flat space analog of sections 3.1.1 and 3.1.2, we can relate the Dirac Green function on the sphere to the Green function $G_S^{(2)}$ of a scalar field of mass $M$.  Defining an auxiliary quantity $\tilde{G}$ as
\begin{equation}
\tilde{G}(z,z^\prime)={\cal D}^\dagger_2 G_S^{(2)}(z,z^\prime)\,,
\end{equation}
we have that 
\begin{equation}
{\cal D}_2 \tilde{G}(z,z^\prime) =\sqrt{g}\, \,\delta^2(z-z^\prime)\,,
\end{equation}
where the scalar and the fermion masses are related by
\begin{equation}
M^2=\frac{1}{4}+m^2\,.
\end{equation}
Thus, the difference \begin{equation}
G_D^{(2)}(z,z^\prime)-\tilde{G}(z,z^\prime)=Q(z,z^\prime)
\end{equation}
 satisfies the Dirac, $
{\cal D} Q(z,z^\prime)=0$, 
 and Helmholtz equation, $(-\Delta_{z}+M^2)Q(z,z^\prime)=0$, without sources.  Therefore it would be identically zero if it where bounded. $Q(z,z^\prime)$ is however unbounded at the extreme points $L_1$ and $L_2$ of the cut. These singularities can be conveniently eliminated by linear combination with the functions $S_i(z)$. Therefore expanding the most general for of $
Q(z,z^\prime)$, imposing the Dirac equation, and using the differential equations for $S_1(z)$ and $S_2(z)$ given in \cite{log} (eqs. (81-84) of that work), one derives the Dirac Green function in terms of quantities related to the scalar one.

Here we need only the part of the trace of $G_D^{(2)}= \tilde{G}+ \, Q$ which is odd in the mass $m$, since the terms even in $m$ do not contribute to the integral (\ref{refe1}). We have
\begin{equation}
  \left. \textrm{tr}\,G^{(2)}_D\,(x,m,a)\right|_{\textrm{odd}}=2 m\, \textrm{tr}\, G_S^{(2)} - \frac{16 \pi a(1-a) m\left(4\, \beta_1\,X_1  \,\cos (x/2)  - b\,B_1 \,{\sin^2 (x)} \right)}{M\,\left( 4\,{{{\beta }_1}}^2 -b^2 \sin^2 (x)  \right)} 
\,.  \label{treta}
\end{equation}
Here $\textrm{tr} \,G_S$, $b$, $\beta_1$, $X_1$ and $B_1$ are functions of $x$ which are the analogous on the sphere to the quantities defined on section 3.1.1 for the plane. They are given in terms of the system of algebraic and ordinary differential equations (\ref{29}-\ref{minuscula}). 

  The result follows combining  (\ref{vein}), (\ref{questa}) and (\ref{refe1}). We have  
\begin{equation}
s_D(x)= \int_0^\infty dt \, \frac{1}{2\,\sinh^2(\pi t)}  \int_{-\infty}^{\infty}dm\,m\,\,\textrm{tr} \,G_D^{(2)}(x,m,-i t)\,.
\end{equation}
The relevant part of $\textrm{tr} \,G_D(x,m,-i t)$ is given by (\ref{treta}) and the formulae at eqs. (\ref{hprima}-\ref{treintaydos}), where we have to make the replacement $a=-i t$. The imaginary part cancels in (\ref{treta}) as it should. 

The function $s_D$ is plotted in figure \ref{ryo} with some lattice results. Table \ref{tatata} shows the coefficients of the Taylor expansion around $x=\pi$ for $s_S$ and $s_D$. Also shown is the coefficient of the term $\sim 1/x$ for $s(x)$ in the limit $x\sim 0$. In this limit the formula (\ref{angulochico}) applies, with $c(t)$ the c-function corresponding to a Dirac field in one dimension.

\begin{table}[t]
\centering
\begin{tabular}{|c|ccccccc|} \hline
 & $c^{(\pi)}_2$ & $c^{(\pi)}_4$ & $c^{(\pi)}_6$ & $c^{(\pi)}_8$ & $c_{-1}^{(0)}$ & $s(\pi/2)$ & $s(3/4 \pi)$   \\  \hline
$s_S$ & $7.81253\, 10^{-3}$ & $5.45402 \,10^{-4}$ & $5.34656 \,10^{-5}$ & $5.40167 \,10^{-6}$ & $7.94 \, 10^{-2}$  & $0.02366$ & $0.005040$\\
$s_D$ & $7.81253 \,10^{-3}$ & $5.01426\, 10^{-4}$ & $4.81299\, 10^{-5}$ & $4.85523\, 10^{-6}$ & $7.22 \, 10^{-2}$ & $0.02329$ & $0.005022$  \\  \hline
\end{tabular} 
\caption{The first four non zero Taylor coefficients of $s_D(x)$ and $s_S(x)$ (a complex scalar)  for $x$ around $\pi$, $s(x)\sim \sum c^{(\pi)}_j\, (x-\pi)^j$,  and the coefficient of the term $1/x$ of these functions for $x\rightarrow 0$, $s(x)\sim  c_{-1}^{(0)}/x$. The value of the functions for $x=\pi/2$ and $x=3/4\, \pi$ are also shown.  The quadratic coefficients of $s_D(x)$ and $s_S(x)$ coincide. }
\label{tatata}
\end{table}

\subsubsection{Lattice Hamiltonian and correlators in two  dimensions}

\noindent 
\textbf{Scalar}

\noindent In the lattice simulations we have taken the following lattice Hamiltonian for a real scalar in three dimensions   
\begin{equation}
{\cal H}=\frac{1}{2}\sum_{n,m=-\infty}^{\infty}( \pi _{n,m}^{2}+(\phi _{n+1,m}-\phi_{n,m})^{2}+(\phi _{n,m+1}-\phi_{n,m})^{2} +m^2 \phi _{n,m}^{2})\,.
\end{equation}
We have set the lattice spacing to one. 
The correlators (\ref{x}) and (\ref{p}) are  
\begin{eqnarray}
\langle \phi_{0,0}\phi_{i,j}\rangle &=&\frac{1}{8\pi^2}\int_{-\pi}^{\pi}dx \int_{-\pi}^{\pi}dy\frac{\cos(i x)\cos(j y)}{\sqrt{2
(1-\cos(x))+2
(1-\cos(y)+m^2)}}\,, \\
\langle \pi_{0,0}\pi_{i,j}\rangle &=& \frac{1}{8\pi^2}\int_{-\pi}^{\pi} dx\int_{-\pi}^{\pi}dy \cos(i x)\cos(j y)}{\sqrt{2
(1-\cos(x))+2
(1-\cos(y))+m^2}\,.
\end{eqnarray}

\noindent
\textbf{Dirac fermion}

\noindent We take the lattice Hamiltonian for a Dirac fermion as
\begin{equation}
{\cal H}=\sum_{n,m} -\frac{i}{2}\left(\left(\Psi^\dagger_{m,n} \gamma^0\gamma^1(\Psi_{m+1,n}-\Psi_{m,n})+\Psi^\dagger_{m,n} \gamma^0\gamma^2(\Psi_{m,n+1}-\Psi_{m,n})\right)-h.c.\right)+m  \Psi_{m,n}^\dagger\gamma^0 \Psi_{m,n}\,. 
\end{equation}
The correlators are
\begin{equation}
\langle \Psi^\dagger_{i,k}\Psi_{j,l}\rangle=\frac{1}{2}\delta_{i,j}\delta_{k,l}+\int_{-\pi}^\pi dx\, \int_{-\pi}^\pi dy\, \frac{m \gamma^0+\sin( x)\gamma^0 \gamma^1+\sin( y)\gamma^0\gamma^2}{8 \pi^2 \sqrt{m^2+\sin^2( x)+\sin^2( y)}} e^{ i (x (i-j)+y (k-l))}\,.
\end{equation}
Due to the fermion doubling on the lattice, one has to divide the lattice results by $4$ in two dimensions in order to get the entropy corresponding to a Dirac field in the continuum limit.  

\smallskip

\noindent
It is relevant to the accuracy of the entropy calculation to evaluate the correlators with enough precision. This can be very time-consuming. We have found it is much faster to evaluate one of the two integrals in the correlators analytically (in terms of polynomials times elliptic functions) using a program for analytic mathematical manipulations, and then doing the last integral numerically.

\subsection{Universal terms in the entanglement entropy for higher dimensions}
\subsubsection{Dimensional reduction}
For free fields, some universal terms in the entanglement entropy in higher dimensions can be obtained from the low dimensional results \cite{boson}. Let us consider sets in $k+d$ spatial dimensions of the form $Z=B\times X$, where $B$ is a box on the first $k$ coordinates $x_1,...x_k$, of sides having lengths $R_i$, $i=1,...,k$, and $X$ is a set in $d$ dimensions. 
We are interested in the entropy of $Z$ in the limit of large $R_i$, which should be extensive in the sides $R_i$. Thus, we can compactify the directions $x_i$, with $i=1,...,k$, by imposing periodic boundary conditions $x_i\equiv x_i+R_i$, without changing the result of the leading extensive term. For a free field of mass $m$ we Fourier decompose it into the corresponding field modes in the compact directions. The problem then reduces to a $d$ dimensional one with an infinity tower of massive fields. For example, for a free scalar we obtain the tower of fields $\phi_{q_1, ..., q_{k}}$, where $q_1$,...,$q_{k}$ are integers, such that the momentum in the first $k$ components is $p_i=\frac{2\pi}{R_i}q_i$. 
From the point of view of the non compact $x_{k+1},..., x_{k+d}$ directions, these fields have masses given by 
\begin{equation}
M_{q_1, ..., q_{k}}^2=m^2+\sum_{i=1}^{k} \left(\frac{2\pi}{R_i}q_i \right) ^2\,. 
\end{equation}
Summing over the contributions of all the decoupled $d$ dimensional fields we have
\begin{equation}
S(Z)=\sum_{\{q_i\}=-\infty}^\infty S(X,M_{q_1, ..., q_{k}})\,.
\end{equation}
In the limit of large $R_i$ we can convert these sums into integrals
\begin{equation}
S(Z)= \frac{k\,{\cal A}}{2^{k}\pi^{k/2}\Gamma(k/2+1)}\int_0^\infty dp\, p^{k-1}\, S(X,\sqrt{m^2+p^2})\,,\label{ala}
\end{equation}
where the transversal area ${\cal A}=\prod_{i=1}^k R_i$.
The universal terms in $S(Z)$ will then come from the ones of $X$ after this integration over the mass. For non scalar fields the factors of the spin multiplicities have to be incorporated into (\ref{ala}).  

One important application of this formula is the case of a thin set $Z$ which is large in all directions excepting one, with size $L$. The leading universal term in the entropy is proportional to the area ${\cal A}$ in the large face transversal to $L$,
 \begin{equation}
\delta S(Z)= -\kappa\, \frac{{\cal A}}{L^{d-1}}\,.\label{tuy}
\end{equation}
The same coefficient $\kappa$ gives the dominant part of the mutual information of two sets $A$ and $B$ with large parallel faces of area ${\cal A}$, at a distance $L$ much smaller than the other dimensions,  
\begin{equation}
I(A,B)\simeq \kappa\, \frac{{\cal A}}{L^{d-1}}\,.\label{tuy1}
\end{equation}

In order to calculate $\kappa$ we use (\ref{ala}) where $X$ is a one dimensional set of length $L$. In consequence $L\,dS(Z)/dL$ is related to the entropic c-function of the one dimensional case. It follows that
\begin{equation}
\kappa(t) =   \frac{ t^{d-1}}{2^{d-2}\,\pi^{\frac{d-1}{2}}\,\Gamma\left(\frac{d-1}{2}\right)}\,\,\int_t^\infty dy_1\, y_1^{-d} \int_0^\infty d y_2\, \, y_2^{d-2}\, c\left(\sqrt{y_1^2+y_2^2}\right)\,,\label{masapan}
\end{equation}
where $t=m L$, and $m$ is the mass of the field in $d$ spatial dimensions.  Here the function $c(t)=(n_B c_B(t)+n_F c_F(t))$, where $n_B$ and $n_F$ are the multiplicity of the bosonic and fermionic degrees of freedom, and $c_B(t)$ and $c_F(t)$ are the one dimensional entropic functions corresponding to a real scalar (boson) and a Majorana fermion fields (see section 3.1.3). In the massless limit $\kappa$ reduces to a constant  
\begin{equation}
\kappa =  \left( (d-1)\,2^{d-2}\,\pi^{\frac{d-1}{2}}\,\Gamma\left(\frac{d-1}{2}\right)\right)^{-1} \int_0^\infty dy \,\,y^{d-2}\, c(y) \,.\label{ttt1}
\end{equation}

The coefficient $\kappa$ can be calculated numerically with good precision. Table (\ref{tatal}) shows $\kappa$ for the first few dimensions. For large dimensions the bosonic and fermionic coefficients approach each other, what is a consequence of the fact the c-function has the same asymptotic large $t$ behavior in both cases, $c(t)\sim \frac{1}{4} t K_1(2 t)$. Using this approximation in (\ref{ttt1}) we have for large $d$
\begin{equation}
\kappa\simeq \frac{\Gamma\left(\frac{d-1}{2}\right)}{2^{d+3}\pi^{(d-1)/2}}\,.
\end{equation} 
This has less than the 3\% error already for $d=5$.

 \begin{table}[t]
\centering
\begin{tabular}{|c|cc|} \hline 
$d$ & $\kappa \textrm{(bosons)}$ & $\kappa \textrm{(fermions)}$ \\  \hline
$2$ & $3.97 \,10^{-2}$& $3.61 \,10^{-2}$ \\  
$3$ & $5.54 \,10^{-3}$ & $5.38 \,10^{-3}$\\ 
$4$ & $1.31 \,10^{-3}$ & $1.30 \,10^{-3}$\\ 
$5$ & $4.08 \,10^{-4}$ & $4.06 \,10^{-4}$\\ \hline
\end{tabular} 
\label{tatal}
\caption{The coefficients $\kappa$ for massless bosons and fermions (one field degree of freedom) for different spatial dimension $d$.}
\end{table}
 
For $d=2$, and considering now $V$ as an angular sector of small angle $x$ instead of a long strip, one should add the contribution of the term (\ref{tuy}) along the length of the angular sector,
 \begin{equation}
\delta S\sim -\kappa \int dL \frac{1}{x L}\sim \frac{\kappa}{x}\log \epsilon\,.\label{ftft}
\end{equation}
 For $d=2$ eq. (\ref{ttt1}) gives $\kappa=\frac{\int_0^\infty dt \,c(t)}{\pi}$, and (\ref{ftft}) gives the logarithmic term in the entropy (\ref{angulochico}) for the small angle limit \cite{log}.  
 
\subsubsection{Two component sets in the large distance limit}
Another interesting universal term is the correction for the entropy of two sets $A$ and $B$ which are far apart \cite{remarks}. This is measured by the mutual information $I(A,B)$, which goes to zero for long separating distance $L$. 
 We find convenient to use the real time approach in order to compute this limit.  
For a free Dirac field we can use the formula (\ref{fas}) for the entropy  
in terms of the resolvent $R(\beta)=\left( C+\beta-\frac{1}{2}\right)^{-1}$ of the field correlator $C$. 
For a massless field we have 
\begin{equation}
C(x,y)=<0|\Psi(x)\Psi(y)^\dagger|0>= \frac{1}{2} \delta(x-y)+i c\, \gamma^i \gamma^0 \frac{(x-y)^i}{|x-y|^{d+1}}\,,
\end{equation}
where $c$ is a constant which depends on $d$, and $\gamma^\mu$ are the Dirac matrices.

Considering $V=A\cup B$ with a large separation vector $\vec{L}$ between $A$ and $B$, the resolvent can be expanded perturbatively,
\begin{equation}
R_V(\beta)=R_V^0(\beta)-R_V^0(\beta){\cal C}_1 R_V^0(\beta)+R_V^0(\beta){\cal C}_1 R_V^0(\beta){\cal C}_1 R_V^0(\beta)-...\,.
\label{financiar1}
\end{equation}
Here $R_V^0(\beta)$ is the unperturbed resolvent, 
\begin{equation}
R_V^0(\beta)=\left(\begin{array}{cc}
R_A(\beta) & 0 \\
0 & R_B(\beta) 
 \end{array}
\right)\,,
\end{equation}
and ${\cal C}_1$ is the field correlator evaluated for the separation vector $\vec{L}$ 
\begin{equation}
{\cal C}_1=i\,c \,\frac{\gamma^i \gamma^0 L^i}{L^{d+1}} \left(\begin{array}{cc}
0& {\mathbb I}_{A,B} \\
-{\mathbb I}_{B,A} & 0 
 \end{array}
\right)\,.
\end{equation}
The kernel ${\mathbb I}_{A,B}(x,y)=1$ for any $x\in A$ and $y\in B$.

The second term in (\ref{financiar}) does not contribute since it has zero trace. The third term is proportional to the squared of the field correlator and leads to 
\begin{equation}
I(A,B)\sim \frac{c^2}{L^{2 d}}\int^\infty_{1/2} d\beta\,\left(\beta-1/2\right)\left[\, \textrm{tr} \left(  \overline{R_{\hat{A}}}(\beta) \overline{R_B^2}(\beta) +  \overline{R_{\hat{B}}}(\beta)  \overline{R_A^2}(\beta)\right)            -\left(\beta\rightarrow -\beta\right)\right]\,,\label{inn}
\end{equation}
where $\hat{V}$ means the set $V$ after an inversion of coordinates followed by a reflection in the plane perpendicular to $\vec{L}$. The bar over the resolvent and the square of the resolvent means sum over the spatial variables, $\overline{{\cal O}}_X=\int_x dx\int_y dy {\cal O}(x,y)$.
Thus, we have the power $I(A,B)\sim L^{-2d}$. The coefficient should have length dimension $2d$, and depend on $A$ and $B$ and their relative position but not on the distance $L$. Moreover, it must be monotonically increasing with $A$ and $B$. For the massive case $I(A,B)$ is also proportional to the field correlator squared and decays exponentially quickly. In particular, in one dimension we have \cite{futuro}
\begin{equation}
I(A,B)\sim \frac{1}{3} \,m^2 \,L^A_t\, L^B_t \,(K_0^2(m L)+ K_1^2(m L)) \,\,\,\,(1 + {\cal O} (L_{A,B}/L)+...)\,,  
\end{equation}
where $L_A^t$ and $L_B^t$ are the total lengths of $A$ and $B$, assumed much smaller than $m^{-1}$ and $L$.

\begin{figure} [tbp]
\centering
\leavevmode
\epsfysize=5.3cm
\bigskip
\epsfbox{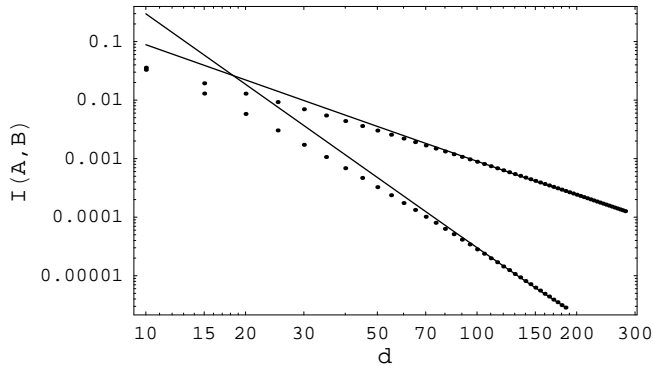}
\caption{The log-log plot of the mutual information $I(A,B)$ for two squares $A$ and $B$ of side 10 lattice points as a function of the separation distance $L$ in two dimensions. The line with larger slope ($I(A,B)\sim~L^{-4}$) corresponds to a Dirac field, while the other to a scalar one ($I(A,B)\sim~L^{-2}$).}
\end{figure}

A similar calculation for scalars on a lattice  shows also a decay going like the field correlator squared. For massless fields this gives $I(A,B)\sim L^{-2(d-1)}$, which is slower than the fermionic case. It is instructive to obtain this behavior with a different approach. The mutual information is an upper bound on correlation \cite{upper}
\begin{equation}
I(A,B)\ge \frac{1}{2}\frac{\left(\left<O_A O_B\right>-\left<O_A\right>\left<O_B\right>\right)^2}{\left\|O_A\right\|^2\left\|O_B\right\|^2}\label{dfg}\,,
\end{equation}
where $O_A$ and $O_B$ are any bounded operators in the local algebras corresponding to $A$ and $B$, with norms $\left\|O_A\right\|$ and $\left\|O_B\right\|$ respectively. We cannot apply this relation directly to the smeared scalar field operator, $\int dx \,\alpha(x) \phi(x)$, with $\alpha(x)$ an smooth function of compact support, since this is not bounded. However, using the inequality (\ref{dfg}) with the unitary operators $O_A=\exp{(i  \int dx \,\alpha(x) \phi(x))}$ and $O_B=\exp(i \int dy \, \beta(y) \phi(x))$, and $\alpha(x)$, $\beta(x)$ vanishing outside $A$, $B$ respectively, we obtain, 
\begin{equation}
I(A,B)\ge \textrm{cons} \left<\phi(0) \phi(\vec{L})\right>^2\,\sim L^{-2(d-1)}.
\end{equation} 

\subsubsection{Logarithmic term for smooth boundaries in four dimensions}
In the Euclidean approach the entanglement entropy is obtained from the effective action $W=-\log(Z)$ calculated in an Euclidean $d+1$ dimensional manifold with conical singularities on $\partial V$.  
We have the general expansion 
\begin{equation}
W=\textrm{non universal} +a_{d+1} \log{\epsilon}+\textrm{finite}\,.\label{polo}
\end{equation}
The non universal terms diverge as inverse powers of the cutoff $\epsilon$. In the case of a free theory $a_{d+1}$ coincides with the corresponding heat kernel coefficient (\ref{ggg}). For a general conformal theory it is determined by the integrated conformal anomaly. In order to see this, consider an infinitesimal rescaling of the metric $g^{\mu\nu}\rightarrow(1-2 \delta \lambda)g^{\mu\nu}$. Since 
\begin{equation}
\frac{2}{\sqrt{g}}\frac{\delta W}{\delta g^{\mu\nu}}=\left<T_{\mu\nu}\right>\,,
\end{equation}
in terms of the energy momentum tensor $T_{\mu\nu}$, we have 
\begin{equation}
\frac{\delta W}{\delta \lambda}=-\int dx^4\, \sqrt{g}  \left<T_{\,\,\,\mu}^\mu\right>\,,
\end{equation}
which is the integrated trace anomaly (classically, for a conformally invariant theory we should have $T_{\,\,\,\mu}^\mu=0$). On the other hand, due to the conformal invariance of the action, scaling the metric as above must give the same result as keeping the metric constant but contracting the cutoff as  $\epsilon\rightarrow (1-\delta \lambda)\epsilon$. Using (\ref{polo}) this gives  
\begin{equation}
a_{d+1}= \int dx^4\, \left<T_{\,\,\,\mu}^\mu\right>\,.
\end{equation}  
The trace anomaly for smooth manifolds has a local expression in terms of polynomials of the curvature tensor of dimensionality equal to $d+1$. Thus, it must vanish for odd spacetime dimensions. In four spacetime dimensions it writes \cite{anomalia}
\begin{eqnarray}
a_4&=& \frac{1}{5760 \pi^2} (a E - 3 c  I)\,,\label{gfgf}\\
E&=&\int dx^4\, (R_{\alpha \beta \mu \nu}R^{\alpha \beta \mu \nu}-4 R_{\mu \nu}R^{\mu \nu}+R^2)\,,\\
I&=& \int dx^4\, (R_{\alpha \beta \mu \nu}R^{\alpha \beta \mu \nu}-2 R_{\mu \nu}R^{\mu \nu}+\frac{1}{3} R^2)\,,\label{last}
\end{eqnarray}
where $a$ and $c$ are the anomaly coefficients, which are a characteristic property of the theory. Here they are normalized such that $a=c=1$ for the real scalar. For a Weyl fermion it is $a=11/2$, $c=3$ \cite{anomalia}.

As they stand these equations cannot be used for evaluating the entanglement entropy in flat space since the bulk curvature tensor $R_{\alpha \beta \mu \nu}=0$, and, on the other hand, there is a conical singularity at $\partial V$. 
 However, it was argued by Fursaev and Solodukhin in \cite{dos} that at lowest order in the deficit angle $\delta=2\pi(1-n)$, the contribution of the surface term is proportional to this deficit,
 \begin{equation}
 a_4\sim (n-1) s\,,\label{yeye}
 \end{equation}
 and $s$ can be calculated using the same expressions (\ref{gfgf}-\ref{last}), with a curvature tensor proportional to a delta-function concentrated in $\partial V$. 
 From (\ref{alpha}) this gives the logarithmic term in the entanglement entropy
 \begin{equation}
 S_{\log}=s \log \epsilon\,.
 \end{equation}
 The value of $s$ in (\ref{yeye}) was obtained in \cite{dos} for the case of zero extrinsic curvature of the surface $\partial V$, and used to calculate corrections to the entanglement entropy of the black hole \cite{dos,cases} (the extrinsic curvature of the horizon vanishes). The contribution of the extrinsic curvature is however crucial for the geometric entropy in flat space. This was obtained recently by Solodukhin in \cite{una}.  His approach consists in demanding conformal invariance of the most general geometric expression for the logarithmic contribution, and then using the holographic anzats of Ryu and Takayanagi (which is applicable to the special case of the entanglement entropy of $SU(N)$ superconformal gauge theories  \cite{hkernel}) in order to calibrate a coefficient. The result in flat space is\footnote{The relation of the present normalization of the anomaly coefficients with the one in \cite{una} is $A=a/(90 \pi^2)$, $B=c/(30 \pi^2)$.} \cite{una}
\begin{equation}
s=\frac{a}{720 \pi} \int_{\partial V}(k_i^{\mu \nu}k^i_{\nu \mu} - k_i^{\mu \mu}k^i_{\mu \mu}) +\frac{c}{240 \pi}   \int_{\partial V}(k_i^{\mu \nu}k^i_{\nu \mu} -\frac{1}{2} k_i^{\mu \mu}k^i_{\mu \mu})\,.\label{turti}
\end{equation}
Here  $k^i_{\mu\nu}=-\gamma^\alpha_\mu \gamma^\beta_\nu \partial_\alpha n^i_\beta$ is the second fundamental form, $n^\mu_i$ with $i=1,2$ are a pair of unit vectors orthogonal to $\partial V$, and $\gamma_{\mu\nu}=\delta_{\mu\nu}-n^i_\mu n^i_\nu$ is the induced metric on the surface. In the first term, the integral $\int_{\partial V}(k_i^{\mu \nu}k^i_{\nu \mu}- k_i^{\mu \mu}k^i_{\mu \mu})=4\pi \chi (\partial V)$, where $\chi (\partial V)$ is the Euler number of the surface, which is purely topological.  

For a sphere (\ref{turti}) gives \cite{una}
\begin{equation}
s=\frac{a}{90}  \,,
\end{equation}
while for a cylinder of length $L$ and radius $R$ it is \cite{una}
\begin{equation}
s= \frac{c}{240}\frac{L}{R}\,.
\end{equation}
The sphere and the cylinder are then sensitive to different anomaly coefficients. 

Note here that for the electromagnetic field it is $a=62$ and $c=12$.  This shows the entanglement entropy of the electromagnetic field cannot be assimilated to the one of a pair of massless scalars. Also note that there are some conflicting results in the literature which do not coincide with the above formulas for the logarithmic coefficient in three spatial dimensions for scalar fields \cite{sch}.

\section{Outlook}
The obvious missing subject on this review about entanglement entropy for free fields is the one of  free gauge fields. In fact, though there are a few works on this sense in the literature, the entanglement entropy for gauge fields has not been fully understood yet. There are some problems in the definition of $S(V)$ in this case, since there is no gauge invariant partition of the Hilbert space as a tensor product of the Hilbert spaces corresponding to degree of freedom on two complementary regions (see for example \cite{gauge}). Perhaps this is the origin of the boundary term noticed in the early works for the electromagnetic field \cite{Kabat}, which would make the area term non-gauge invariant. At present it is not known if the mutual information is well defined in the presence of gauge symmetries. 

Other problems directly related to free fields also seem to be worth of further work. One is the problem of defining the real time formulation for bosons directly in the continuum. In this case it is easy to see by dimensionality arguments that the correlators $X$ and $P$ of section 2.2.1 are not well defined in the continuum, while $XP$, which is the operator relevant to the entropy calculation, has the right units. One should be able to obtain this continuum operator directly, without passing through a regularization of $X$ and $P$. We think this issue has to be understood before attacking the problem posed by the localization of Abelian gauge fields. 

Other interesting research subjects include the treatment of excitations, which seem to allow a systematic formulation in the real time approach, and the development of a perturbative theory for the reduced density matrix.

 \section*{Acknowledgments}
We thank P. Calabrese, J. Cardy and B. Doyon for their kind invitation to collaborate with this review to a special issue on {\sl Entanglement entropy in extended quantum systems} in Journal of Physica A. 
 We also thank Cesar Fosco for many fruitful discussions. This work was financially supported by CONICET, ANPCyT and UNCuyo.

\end{document}